%% file: rp.tex
\documentclass[twoside,leqno,twocolumn]{article}

\usepackage[letterpaper]{geometry}

\usepackage{ltexpprt}
\usepackage{hyperref}
\usepackage{multirow}
\usepackage{comment}
\usepackage{graphicx}
\usepackage{graphics}
\usepackage[section]{placeins} 
\usepackage{svg} 
\usepackage{amsmath,amssymb,amscd}
\usepackage{cleveref}
\usepackage{algorithm}
\usepackage{algorithmic}
\usepackage[algo2e,ruled,vlined,linesnumbered]{algorithm2e}
\SetAlFnt{\small}
\SetAlCapFnt{\small}
\SetAlCapNameFnt{\small}
\algsetup{linenosize=\tiny}
\usepackage{calc}

\usepackage{soul}

\input{tikz-includes}

\def\bx{\mathbf{x}}   
\def\lim{{\xi}}    

\def\ra{$r-$adaptivity}

\def\bbx{\bar{\bx}}

\def\bw{\bar{w}}

\def\ws{$w_{\sigma}$}
\def\erft{$e_{\tt ref}$}

\def\delp{$\Delta p$}
\def\delpr{$\Delta p_{\tt ref}$}

\newcommand\anon[2]{{#2}}

\begin{document}

\title{
Mixed-Order Meshes through $rp$-adaptivity for
Surface Fitting to Implicit Geometries\thanks{
Performed under the auspices of the U.S. Department
of Energy under Contract DE-AC52-07NA27344 (LLNL-CONF-854941).
}
}

\anon{
    \author{Submission ID 1004}
}
{ \author{
Ketan Mittal\thanks{Lawrence Livermore National Laboratory, Livermore, CA, U.S.A. \{mittal3,dobrev1,kolev1,tomov2\}@llnl.gov.} \and \hspace{-10mm} \thanks{Corresponding Author.} \and
Veselin A. Dobrev\footnotemark[2] \and
Patrick Knupp\thanks{Dihedral LLC, Bozeman, MT, U.S.A. knupp.patrick@gmail.com.} \and
Tzanio Kolev\footnotemark[2] \and
Franck Ledoux\thanks{CEA, DAM, DIF, F-91297, Arpajon, France. franck.ledoux@cea.fr.} \and
Claire Roche\thanks{CEA-CESTA, Le Barp, France. claire.roche@cea.fr.} \and
Vladimir Z. Tomov\footnotemark[2]
}}

\date{}

\maketitle


\fancyfoot[R]{\scriptsize{Copyright \textcopyright\ 2024 by SIAM\\
Unauthorized reproduction of this article is prohibited}}





\begin{abstract} \small\baselineskip=9pt
Computational analysis with the finite element method requires geometrically
accurate meshes. It is well known that high-order meshes can accurately capture
curved surfaces with fewer degrees of freedom in comparison to low-order
meshes. Existing techniques for high-order mesh generation typically output
meshes with same polynomial order for all elements. However, high order
elements away from curvilinear boundaries or interfaces increase the
computational cost of the simulation without increasing geometric accuracy.
In prior work \cite{barrera2023high,TMOP2021IMR}, we have presented one such approach for generating body-fitted \emph{uniform-order} meshes that takes a given mesh and morphs it to align with the surface of interest prescribed as the zero
isocontour of a level-set function.  We extend this method to generate
mixed-order meshes such that curved surfaces of the domain are discretized with
high-order elements, while low-order elements are used elsewhere. Numerical
experiments demonstrate the robustness of the approach and show that it can be
used to generate mixed-order meshes that are much more efficient than high
uniform-order meshes. The proposed approach is purely algebraic, and extends to
different types of elements (quadrilaterals/triangles/tetrahedron/hexahedra) in
two- and three-dimensions.
\anon{Authors are anonymized for double-blind review.}{}
\end{abstract}


\section{Introduction}

Meshes are an integral part of computational analysis using techniques such as the finite element method (FEM) and the spectral element method (SEM). From a mesh generation perspective, there are broadly two main requirements. First, the mesh must accurately capture
the domain boundaries and multimaterial interfaces. Second, the elements of the mesh must have good quality in terms of shape, size, and orientation to ensure accuracy in the numerical solution of the PDE of interest.
This latter issue of mesh quality has been studied extensively, and there are various techniques based on $r$- and $h$-adaptivity for improving the quality of an existing mesh and increasing the accuracy of the solution \cite{alauzet2016decade,aparicio2023combining,frey2005anisotropic,fidkowski2011review,loseille2010fully,marcon2019variational,odier2021mesh,yano2012optimization,zhang2018curvilinear}.
The first issue of automatically generating a geometrically accurate mesh
is a challenging open-question, and has led to the development of techniques
that support meshes that are not aligned with the exact geometry \cite{ingram2003developments,main2018shifted,mittal2005immersed,schneiders2013accurate,peskin2002immersed}.
These methods usually require complex numerical techniques to ensure robustness
and accuracy, and classical approaches based on body-fitted/domain-conforming
meshes are usually preferred in many situations.

In the context of body-fitted meshes, it is well known that a high-order mesh can accurately capture curved surfaces at a lower computational cost in comparison to its lower order counterpart. From a mesh generation point of view, it is convenient to generate a uniform-order mesh, i.e. a mesh in which all the elements are represented by the same number of degrees of freedom.
Uniform-order meshes are also necessary when the FEM framework cannot support mixed-order/$p$-refined/$p$-adaptive meshes.
For frameworks that support $p$-refined meshes, high-order elements can increase the computational cost without increasing accuracy when they are used in region where they are not necessary, e.g., around planar surfaces.
Note that mesh curvature can also be dictated by a discrete simulation field,
e.g., in Arbitrary Lagrangian Eulerian (ALE) simulations the mesh adaptivity
is usually driven by the numerical solution \cite{TMOP2020CAF, TMOP2021EWC}.
For the purposes of this work we focus on geometry-driven mesh curvature.

From a historical perspective, body-fitted high-order meshes are typically generated by starting with a linear body-fitted mesh, elevating the mesh to a higher polynomial order, and projecting the higher order nodes of the boundary elements to the domain boundary  \cite{xie2013generation,fortunato2016high,moxey2015isoparametric,gargallo2016distortion,toulorge2016optimizing,poya2016unified,ruiz2022automatic,marcon2020rp}. Each of these approaches deals with uniform-order meshes.
The literature for $p$-refined mesh generation is very sparse,
except mainly the work by Karman et al. \cite{karman2022mixed}.
Therein, the authors present a method for taking a CAD geometry and a
corresponding linear body-fitted mesh, and sequentially elevating the order of the elements at the boundary, blending the polynomial order increase in the interior, and smoothing the mesh to ensure mesh validity.
The $p$-refinement/order-elevation for the boundary elements is based on the error between the discretized surface and the actual surface,
while the order of interior elements is determined by measuring the difference
in the shapes of adjacent elements with different polynomial orders.

We explore mesh $p$-adaptivity in the case when the target surface is
described by the zero isocontour of a level-set function.
Level-set based descriptions are commonly
used for material interfaces in multimaterial configurations \cite{Osher1994}, and evolving geometries
in shape and topology optimization
\cite{sokolowski1992introduction,allaire2014shape,hojjat2014vertex}, amongst other applications.
In Prior work \cite{TMOP2021IMR,barrera2023high}, we have presented a technique for
taking any mesh (e.g., uniform Cartesian-aligned mesh) and morphing it using
node movement ($r$-adaptivity) to align with the target surface.
This approach has proven to be robust and efficient at obtaining
body-fitted high-order meshes for shape optimization applications in
the FEM framework \cite{barrera2023high}.

The main contribution of this work is to extend the $r$-adaptivity
framework with $p$-refinement to produce mixed-order meshes.
The proposed approach differs from Karman et al. in several ways,
namely, \cite{karman2022mixed} addresses $p$-adaptivity in the context of classical mesh generation, whereas we pose the problem as mesh adaptivity.
The motivation for this approach is to be able to use existing FEM frameworks
for obtaining body fitted meshes without
having to couple with specialized mesh generation software.
Consequently, our method assumes that the surface is prescribed by a discrete level-set function, instead of being given through CAD.
Furthermore, our $r$-adaptation approach aligns the mesh with the zero
level-set, while simultaneously ensuring good mesh quality
through minimization of a global objective.
In contrast, \cite{karman2022mixed}
uses a sequence of nodal projections (mesh alignment) followed by mesh smoothing to ensure mesh validity. Finally,
instead of using the distance between the actual surface and the discretized surface for determining $p$-refinements,
we use a level-set function based estimator.

The remainder of the paper is organized as follows. We introduce the main components of the $rp$-adaptivity framework in Section \ref{sec_prelim}.
Next, in Section \ref{sec_method} we go into the deeper technical
details of the $p$-adaptive mesh alignment approach.
Finally, we present various numerical experiments to demonstrate the efficiency of mixed-order meshes
in comparison to uniform-order meshes in Section \ref{sec_results}.
Summary and directions for future work are given in Section \ref{sec_concl}.


\section{Preliminaries}
\label{sec_prelim}

In this section, we discuss the key components of our finite element framework that are essential for understanding
the new $rp$-adaptivity method. We first introduce some mathematical notation relevant to the method.
Next, we describe how $p$-adaptivity constraints are imposed for degrees of freedom between elements of
different polynomial order, and finally discuss the target matrix optimization paradigm (TMOP)-based
framework for mesh alignment using \ra.

\subsection{Discrete Mesh Representation}
\label{subsec_mesh}

In our finite element based framework, the domain $\Omega \subset \mathbb{R}^d$
is discretized as a union of curved mesh elements, $\Omega^e$, $e=1\dots N_E$,
each of order $p$.  To obtain a
discrete representation of these elements, we select a set of scalar basis
functions $\{ \bw_i \}_{i=1}^{N_p}$, on the reference element $\bar{\Omega}^e$.
In the case of tensor-product elements (quadrilaterals in 2D and hexahedra in 3D),
$N_p = (p+1)^d$, and the basis spans the space
of all polynomials of degree at most $p$ in each variable.
These $p$th-order basis functions are typically chosen to be Lagrange
interpolants at the Gauss-Lobatto nodes of the reference element.
The position of an
element $\Omega^e$  in the mesh $\mathcal{M}$ is fully described by a matrix
$\bx_e$ of size $d \times N_p$ whose columns represent the coordinates
of the element {\em degrees of freedom} (DOFs).
Given $\bx_e$, we introduce the map between
the reference and physical element, $\Phi_e:\bar{\Omega}^e \to \mathbb{R}^d$:
\begin{equation}
\label{eq_x}
\bx(\bbx) =
   \Phi_e(\bbx) \equiv
   \sum_{i=1}^{N_p} \mathbf{x}_{e,i} \bw_i(\bar{\bx}),
   \,\, \bbx \in \bar{\Omega}^e, ~ \bx \in \Omega^e,
\end{equation}
where $\mathbf{x}_{e,i}$ denotes the $i$-th column of $\bx_e$, i.e.,
the $i$-th node of element $\Omega^e$.

Throughout the manuscript, $\bx$ will denote the position function defined
by \eqref{eq_x}, $\bx_e$ will denote the element-wise vector/matrix of nodal locations for element $\Omega^e$,
and $\bx_E=[\bx_1, \bx_2, \dots \bx_{N_E}]$ will denote the global vector of nodal locations for all elements.

\subsection{TMOP for Mesh Quality Improvement via \ra}\label{subsec_tmop_ra}

For a given element with nodal coordinates $\bx_e$, the Jacobian
of the mapping $\Phi_e$ at any reference point $\bbx \in \bar{\Omega}^e$ is
\begin{equation}
\label{eq_A}
  A_{ab}(\bbx) = \frac{\partial x_{a}(\bbx)}{\partial \bar{x}_b} =
    \sum_{i=1}^{N_p} x_{i,a} \frac{\partial \bar{w}_i(\bbx)}{\partial \bar{x}_b}, \quad a,b = 1 \dots d
\end{equation}
where $x_a$, represents the $a$th component of $\bx$ \eqref{eq_x}, and $x_{i,a}$ represents the $a$th component of $\bx_{e,i}$, i.e., the $i$th DOF in element $\Omega^e$.
The Jacobian matrix $A$ represents the local deformation of the physical element $\Omega^e$ with respect to the
reference element $\bar{\Omega}^e$ at the reference point $\bbx$. This matrix plays an important
role in FEM as it is used to determine mesh validity ($det(A)$ must be greater than 0 at every point in the mesh),
and also used to compute derivatives and integrals. The Jacobian matrix can ultimately impact the accuracy
and computational cost of the solution \cite{mittal2019mesh}.
This lends to the central idea of TMOP of optimizing the mesh to control the
local Jacobian $A_{d \times d}$ in the mesh.

The first step for mesh optimization with TMOP is to specify a target transformation matrix $W_{d \times d}$, analogous to $A_{d \times d}$, for each point in the mesh. Target construction is typically guided by the fact that any Jacobian matrix can be written as a composition of four geometric components \cite{knupp2022geometric}, namely volume, rotation, skewness, and aspect-ratio:
\begin{equation}
\label{eq_W}
W_{d\times d} = \underbrace{\zeta}_{\text{[volume]}} \circ
    \underbrace{R_{d\times d}}_{\text{[rotation]}} \circ
    \underbrace{Q_{d\times d}}_{\text{[skewness]}} \circ
    \underbrace{D_{d\times d}}_{\text{[aspect ratio]}}.
\end{equation}
In practice, the user may specify $W$ as a combination of any of the four fundamental components that they are interested in optimizing the mesh for. For the purposes of this paper, we are mainly concerned with ensuring good element \emph{shape} (skewness and aspect-ratio), so we set $W$ to be that of an ideal element, i.e.,
square for quad elements, cube for hex elements, and
equilateral simplex for triangles and tetrahedrons.
Advanced techniques on how $W$ can be constructed for optimizing different geometric parameters, or even for automatically adapting the mesh to the solution of the PDE
are given in \cite{TMOP2020CAF,TMOP2021EWC,knupp2019target}.

\begin{figure}[tb!]
\centerline{\includegraphics[width=0.3\textwidth]{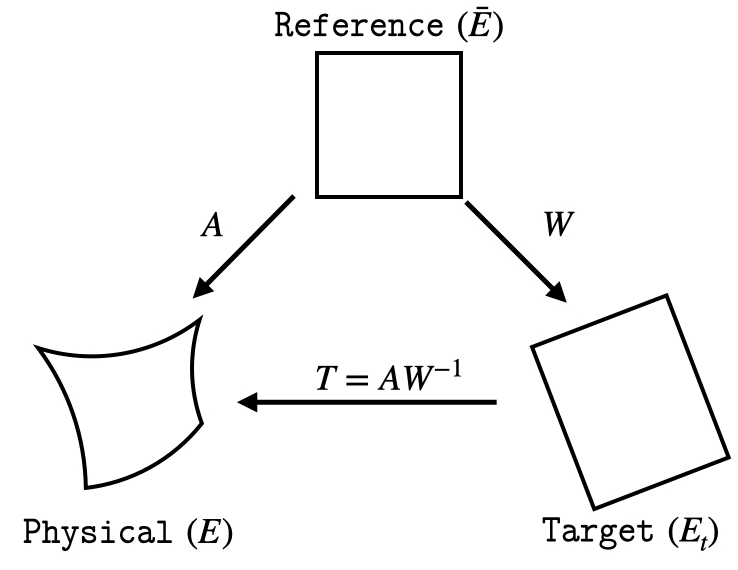}}
\vspace{-1mm}
\caption{Schematic representation of the major TMOP matrices.}
\label{fig_tmop}
\vspace{-3mm}
\end{figure}

The next key component in the TMOP-based approach is a mesh quality
metric that measures the deviation between the current Jacobian transformation $A$
and the target transformation $W$.
The mesh quality metric $\mu(T)$, $T = A W^{-1}$ in Figure \ref{fig_tmop},
compares $A$ and $W$ in terms of some of the geometric parameters.
For example, $\mu_{2,s}=\frac{\mid T \mid^2}{2\tau}-1$ is a \emph{shape} metric\footnote{The metric subscript follows the numbering in \cite{Knupp2020, knupp2022geometric}.}
that depends on the skewness and aspect ratio components, but
is invariant to orientation/rotation and volume. Here, $|T|$ and $\tau$ are the Frobenius norm and determinant of $T$, respectively. Similarly,
$\mu_{77,v}=\frac{1}{2}(\tau-\tau^{-1})^2$ is a
\emph{size/volume} metric that depends only on the volume of the element.
We also use \emph{shape}$+$\emph{size} metrics such as $\mu_{80,vs} =\gamma \mu_{2,s} + (1-\gamma) \mu_{77,v}$,
$0 \leq \gamma \leq 1$, that depend on volume, skewness and aspect ratio,
but are invariant to rotation.

\begin{figure}[t!]
\begin{center}
$\begin{array}{cc}
\includegraphics[height=0.2\textwidth]{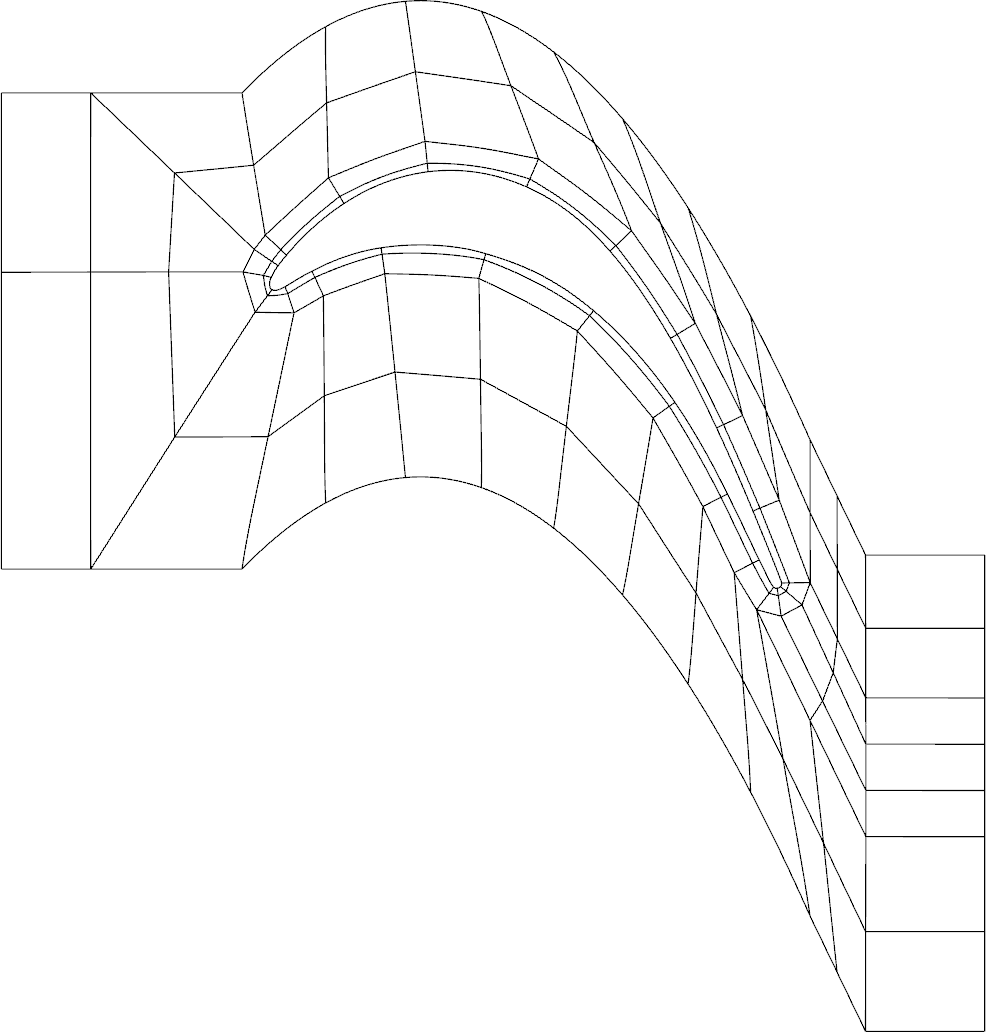} &
\includegraphics[height=0.2\textwidth]{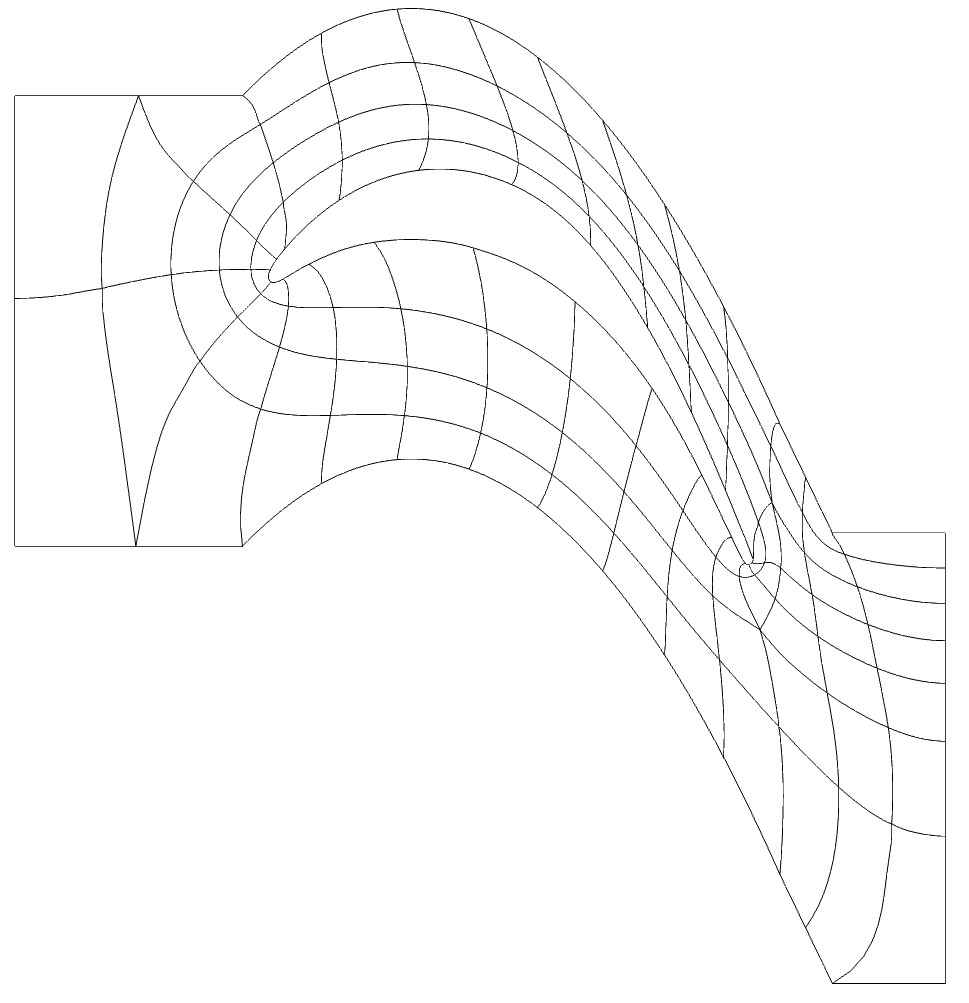}
\vspace{-3mm} \\
\textrm{(a)} & \textrm{(b)} \\
\end{array}$
\end{center}
\vspace{-1mm}
\caption{(a) Original and (b) optimized 4th order mesh for a turbine blade.}
\label{fig_blade}
\vspace{-3mm}
\end{figure}

Using the mesh quality metric, the mesh optimization problem is minimizing the
global objective:
\begin{equation}
\label{eq_F_full}
    F(\bx) = \sum_{\Omega^e \in \mathcal{M}} \int_{\Omega^{e_t}} \mu(T(\bx)) d\bx_t,
\end{equation}
where $F$ is a sum of the TMOP objective function for each element in the mesh,
and $\Omega^{e_t}$ is the target element corresponding to the element $\Omega^e$.
Minimizing \eqref{eq_F_full} results in node movement such that
the local Jacobian transformation $A$ resembles the target transformation $W$ as close
as possible at each point, in terms of the geometric parameters enforced by the mesh quality metric.
Figure \ref{fig_blade} shows an example of high-order mesh optimization for a turbine blade using $W=I$ with a shape metric. The resulting optimized mesh has elements closer to unity aspect ratio and skewness closer to $\pi/2$ radians in comparison to the original mesh, as prescribed by the target $W=I$.

\subsection{$r$-adaptivity for Mesh Quality Improvement and Surface Alignment}\label{subsec_tmop_plus_fitting}

\begin{figure}[tb!]
\begin{center}
$\begin{array}{cc}
\includegraphics[height=0.2\textwidth]{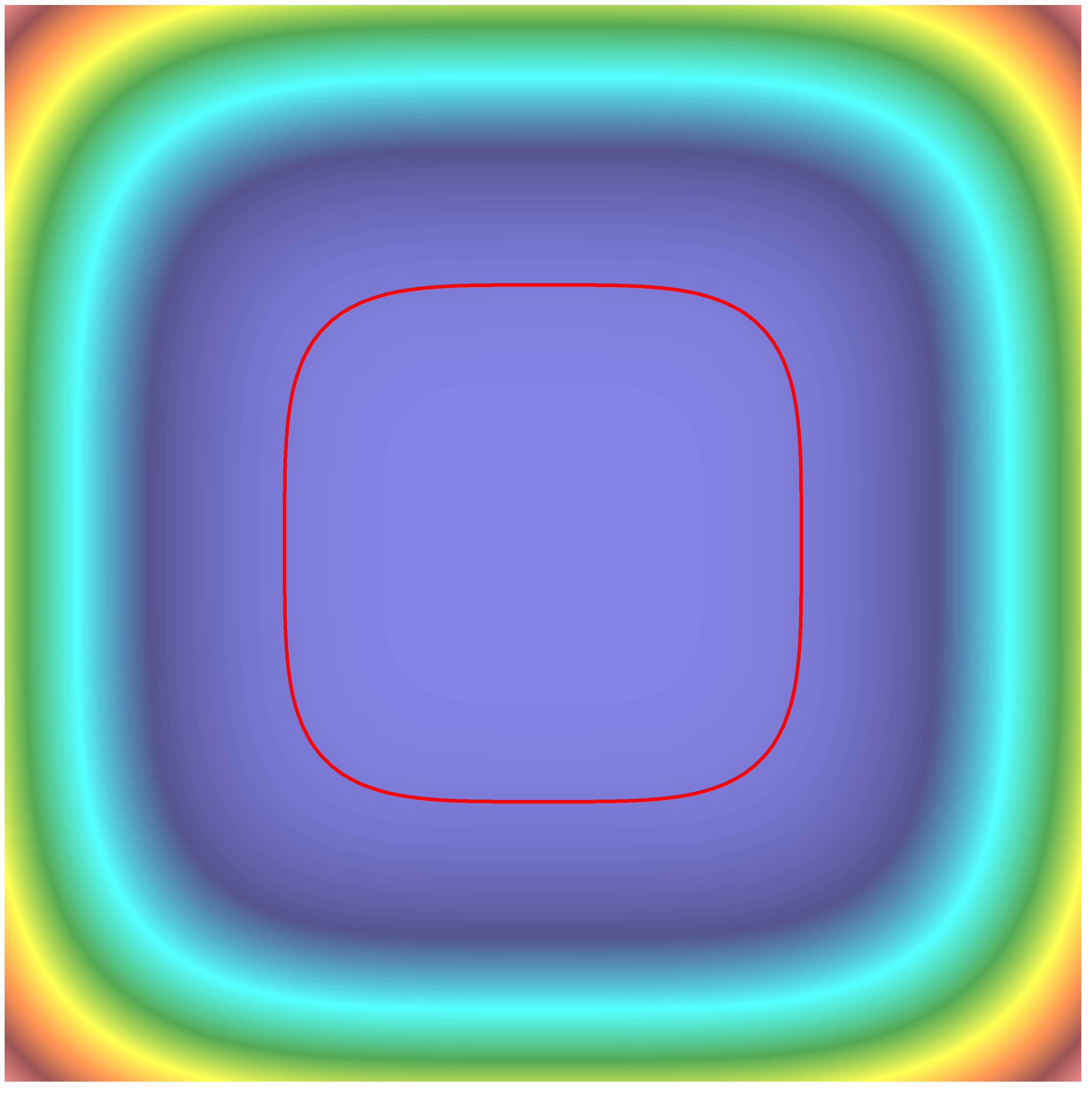} &
\includegraphics[height=0.2\textwidth]{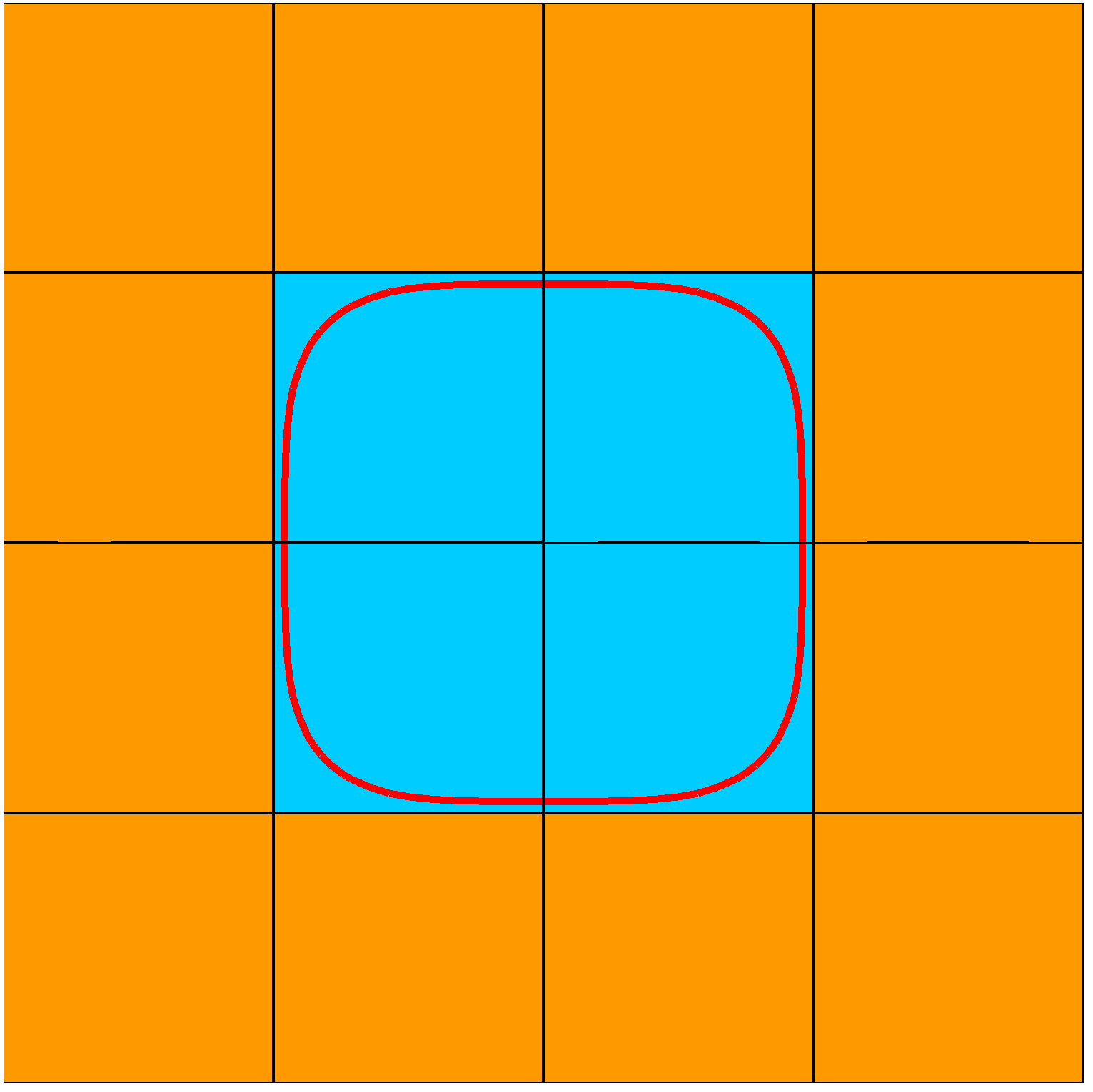} \\
\textrm{(a) $\sigma(\bx)$ \& 0 isocurve.} &
\textrm{(b) Initial mesh, $N_E=16$.} \\
\includegraphics[height=0.2\textwidth]{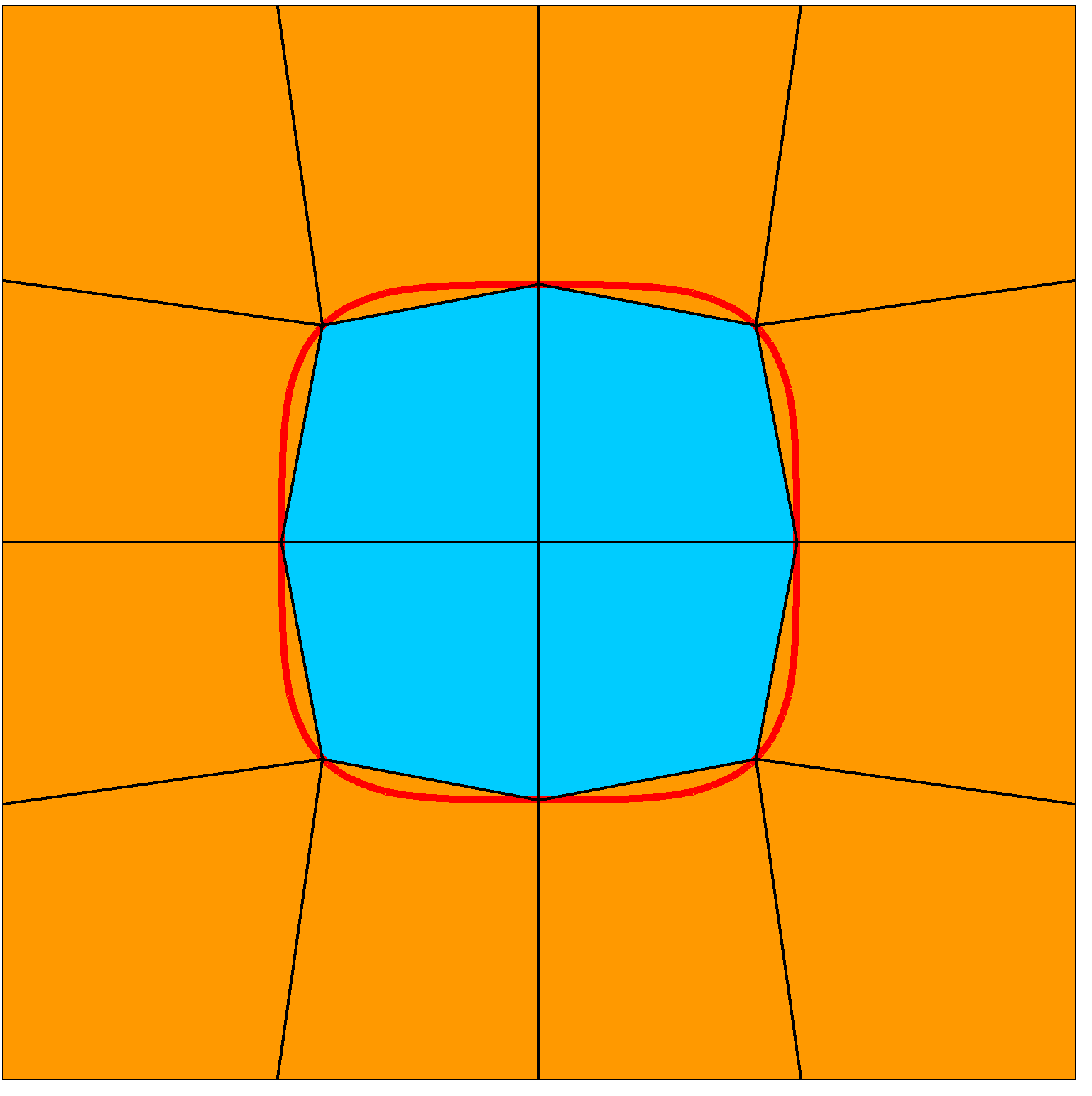} &
\includegraphics[height=0.2\textwidth]{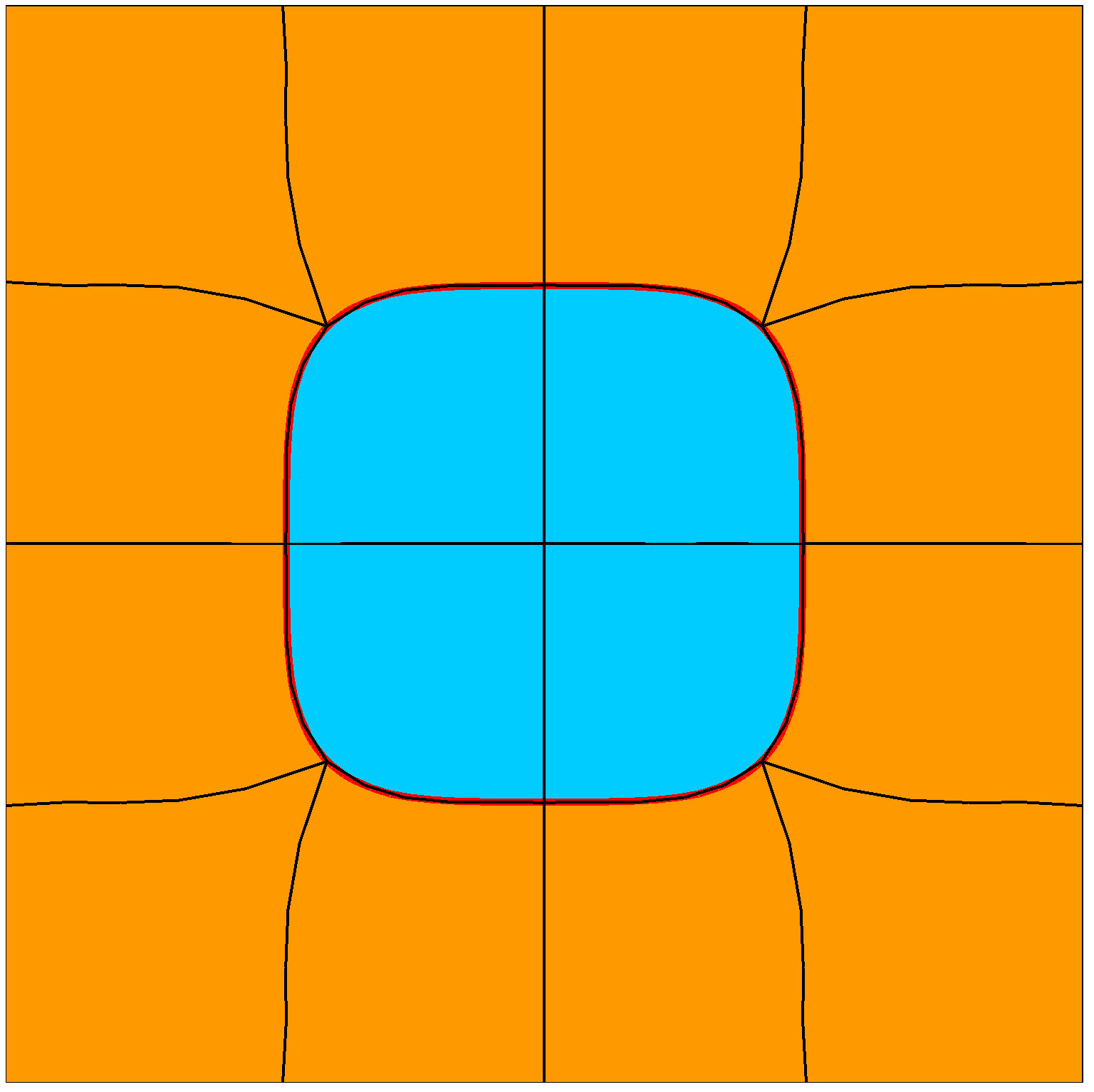} \\
\textrm{(c) Optimized, $p=1$.} &
\textrm{(d) Optimized, $p=3$.} \\
\includegraphics[height=0.2\textwidth]{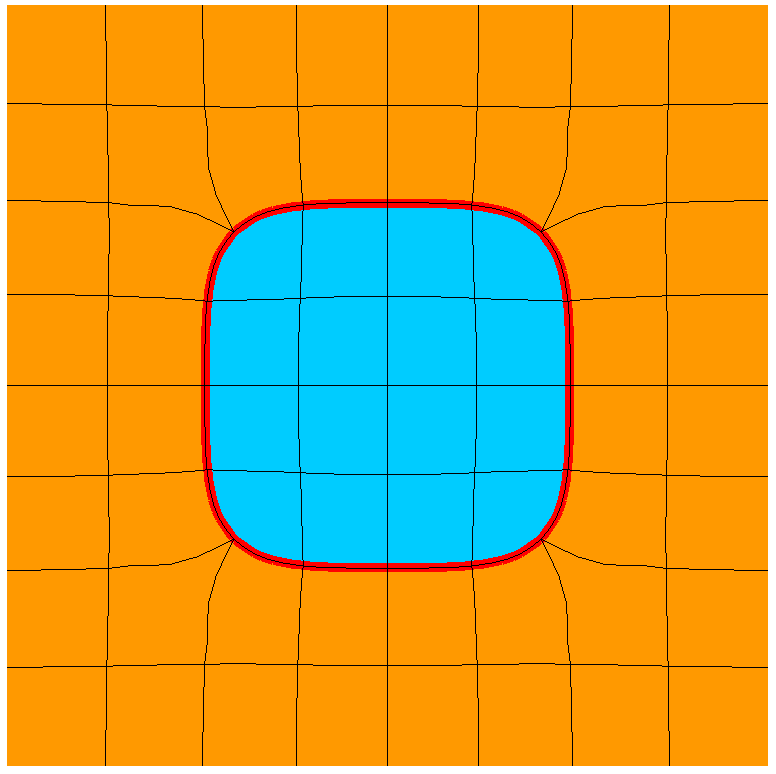} &
\includegraphics[height=0.2\textwidth]{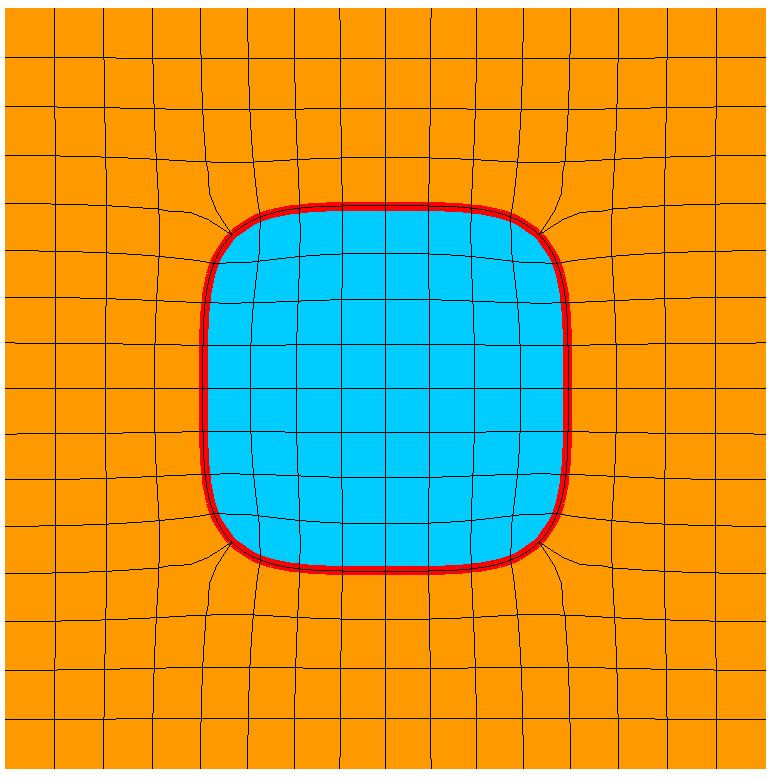} \\
\textrm{(e) $N_E=64, p=3$.} &
\textrm{(f) $N_E=256, p=3$.} \\
\end{array}$
\end{center}
\vspace{-5mm}
\caption{(a) Level set function $\sigma(\bx)$ with zero isocontour in red,
(b) a Cartesian mesh with material interface nodes to be aligned to the zero level set of $\sigma(\bx)$.
(c) Linear mesh and (d) cubic mesh optimized to align with the level-set. Uniformly refined and optimized cubic meshes
are also shown with (e) $N_E=64$ and (f) $N_E=256$ elements.}
\vspace{-3mm}
\label{fig_quad_squircle}
\end{figure}

In the framework of mesh morphing to obtain body-fitted meshes,
it is assumed that the target surface is described as the isocontour
of a smooth discrete level-set function, $\sigma(\bx)$.
Figures \ref{fig_quad_squircle}(a) and (b) show a simple example of a \emph{squircle} interface
represented by the zero isocontour of the level set function,
$\sigma(\bx) = (x-0.5)^4 + (y-0.5)^4 - 0.24^4$,
on a third-order multimaterial quadrilateral mesh.
To effect alignment of the multimaterial interface with $\sigma(\bx)=0$,
the TMOP objective function \eqref{eq_F_full} is augmented as follows:
\begin{equation}
\label{eq_F_full_sigma}
  F(\bx) = \underbrace{\sum_{E \in \mathcal{M}} \int_{E_t}
  \mu(T(\bx)) d\bx_t}_{F_{\mu}} +
  \underbrace{w_{\sigma} \sum_{s \in S} \sigma^2(\bx_s)}_{F_{\sigma}}.
\end{equation}
Here, $F_{\sigma}$ is a penalty-type term that depends on the penalization
weight \ws, the set of nodes $\mathcal{S}$ discretizing the set $\mathcal{F}$
of mesh faces/edges to be aligned to the level set,
and the level set function $\sigma(\bx)$, evaluated at the positions $\bx_s$
of all nodes $s \in \mathcal{S} \in \mathcal{F}$.
In Figure \ref{fig_quad_squircle}(b), $\mathcal{F}$ is thus the union of
all faces between the two materials (colored blue and orange),
and $\mathcal{S}$ is the set of all the nodes that are located on these faces.

Minimizing $F_{\sigma}$ represents weak enforcement of $\sigma(\bx_s) = 0$,
only for the nodes in $\mathcal{S}$, while ignoring the values
of $\sigma$ for the nodes outside $\mathcal{S}$.
Minimizing the full nonlinear objective function, $F = F_{\mu} + F_{\sigma}$,
produces a balance between mesh quality and surface fitting.

The objective function \eqref{eq_F_full_sigma} is usually minimized
by the Newton's method.
This requires the first- and second-derivative of the objective ($F_\mu$ and $F_{\sigma}$) with respect
to the node position.
The Newton iterations are typically done until the maximum fitting error $|\sigma|_{\mathcal{S},\infty}$
is below a user-specified threshold:
\begin{eqnarray}
\label{eq_max_fit_err}
|\sigma|_{\mathcal{S},\infty} := \max_{s \in {\mathcal{S}}} |\sigma(x_s)|.
\end{eqnarray}
Here, the maximum fitting error is the maximum value of the level-set function evaluated at the
nodes $s \in \mathcal{S} \in \mathcal{F}$.
Further implementation details are provided in \cite{barrera2023high,TMOP2021IMR}.
Figures \ref{fig_quad_squircle}(c) and (d) show a first- and third-order optimized $4\times4$ mesh, respectively,
obtained by minimizing \eqref{eq_F_full_sigma}.
Figures \ref{fig_quad_squircle}(e) and (f)
show optimized cubic $8\times8$ and $16\times16$ meshes, respectively.

The level-set in Figure \ref{fig_quad_squircle} is representative of why $p$-refined
meshes are important. Here, the squircle level-set has regions of both
high curvature (resembling a circle) and low curvature (resembling a square).
Using a linear mesh, as in Figure \ref{fig_quad_squircle}(c), results in an interface aligned
mesh where the element vertices are located exactly on the zero level-set.
This is due to the node-wise formulation \eqref{eq_F_full_sigma}.
But in a continuous sense, it is clear that the linear mesh does not align well.
This can be addressed by using a high-order mesh as shown in
Figure \ref{fig_quad_squircle}(d), which aligns more closely with the zero-isocontour, even in the regions of high curvature.
In Figures \ref{fig_quad_squircle}(d)-(f), we also see that as the
mesh is uniformly refined, there are a lot more high-order elements than needed, both in the
regions of low curvature on the interface and away from the interface.
Such extraneous high-order elements could be avoided by using $p$-refined meshes,
which we will address in the next section.

While it is assumed that the prescribed level-set function $\sigma(\bx)$
is smooth, this is often not the case for real-world applications.
The node-wise formulation in \eqref{eq_F_full_sigma} has proven to
be more robust in comparison to an integral-based formulation
when the level-set function is not smooth.
For smooth functions, an integral-based formulation is expected to be better
suited for $p$-adaptivity, and we will explore this in future work.

\subsection{$p$-Adaptivity Constraints} \label{subsec_fem_pref}
\begin{figure}[tb!]
    \centering
    \adjustbox{width=0.9\linewidth,valign=b}{\input{./figures/pconstraint}}
    \vspace{-3mm}
    \caption{Schematic showing the DOFs for a $p$-refined mesh.}
    \label{fig_p_constraint}
    \vspace{-3mm}
\end{figure}
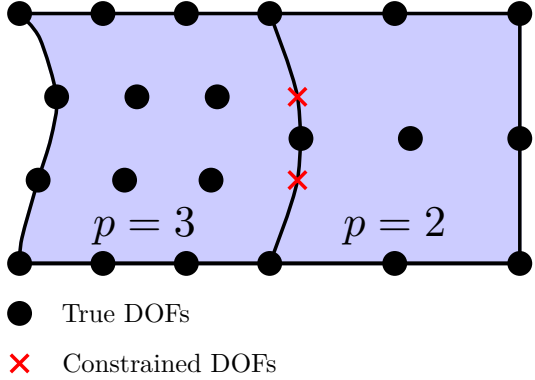

$p$-refinement introduces hanging/non-conforming nodes between elements of
different polynomial orders. In our framework, the hanging nodes on a shared edge/face from higher-order element interpolate from the nodes of the
lower-order element. Thus, the accuracy of a discrete finite element function along a shared edge/face is limited by the lowest polynomial order of adjacent elements.
Consequently, when we $p$-refine for mesh fitting at a certain mesh face,
the polynomial order of the elements on both sides of the face is elevated.
Figure \ref{fig_p_constraint} shows a simple example of a two element $p$-refined mesh with $p=2$ and 3 depicting the \emph{constrained} and \emph{true} DOFs. A detailed description of how general finite element assembly is handled with hanging nodes in our framework is provided in \cite{MFEM2021,cerveny2019nonconforming}.

\section{Methodology}
\label{sec_method}
In this section, we describe our approach for $p$-adaptivity.

\subsection{Level-Set Function Representation and Error Computation for Mesh Fitting}\label{sec_function_rep}

In our framework, we allow the level-set function to be represented on a different background mesh $\mathcal{M}_B$ than
the mesh to be morphed, i.e., $\sigma_B(\bx_B)$, where $\bx_B$ represents the positions of the background mesh.
This decoupling is especially critical when the current mesh $\mathcal{M}$
does not have enough resolution to represent the level-set function
and its gradients with sufficient accuracy near the zero level set.
Since the discretization error in representing a finite element function depends on the element size $h$
and the polynomial order $p$, we typically use a
background mesh that is adaptively refined around the zero isocontour
of the level-set function and has sufficiently high polynomial order.

The example in Figure \ref{fig_quad_squircle} demonstrates that a nodewise approach is not sufficient for measuring the
accuracy of mesh alignment to the level-set. We instead measure the alignment error on each face $f \in \mathcal{F}$ as the squared $\mathcal{L}^2$ norm
of the level-set function on that face:
\begin{eqnarray}
\label{eq_int_fit_err}
e_{f \in \mathcal{F}} := ||\sigma_B||_{\mathcal{L}^2,f}^2 = \int_{f} \sigma_B^2(\bx).
\end{eqnarray}
Here we compute the integrated fitting error for a face/edge of the mesh to be
morphed ($f \in \mathcal{F} \in \mathcal{M}$) with
respect to the level-set function $\sigma_B$ defined on the $\mathcal{M}_B$.
This integration requires interpolation of $\sigma_B$ at the quadrature points
associated with $f$.

\subsection{Level-Set Interpolation from $\mathcal{M}_B$ to $\mathcal{M}$}
\label{subsec_remap}

Since we use a background mesh $\mathcal{M}_B$ for $\sigma_B(\bx_B)$,
it is required to transfer the
level set function and its derivatives from $\mathcal{M}_B$ to the
nodes $\mathcal{S} \in \mathcal{M}$ prior to each Newton iteration,
and at the quadrature points associated with $f \in \mathcal{F}$ prior
to computing the integrated error \eqref{eq_int_fit_err}.
This transfer between the background mesh and
the current mesh is done using \emph{findpts}, a high-order interpolation
library \cite{gslibrepo}.
The \emph{findpts} library enables high-order interpolation at arbitrary points using a sequence of three functions.
First, in a pre-processing step ({\tt findpts\_setup}),
\emph{findpts} constructs some internal data structures based on the input mesh that allow it to
do a fast parallel search for arbitrary points in the second step.
Next, the {\tt findpts} function takes as an input a set of points
$\bx^* = (\bx_1^*,\bx_2^*\dots \bx_b^*)$, where $b$ is the number of points to be found,
and determines the \emph{computational} coordinates of each point.
These computational coordinates are $\mathbf{q}_j^* = (e^*,p^*,\bar{x}^*,\bar{y}^*,\bar{z}^*)_j$ the element $e^*$,
processor $p^*$, and the corresponding reference space coordinates $\bar{\bx}^*=(\bar{x}^*,\bar{y}^*,\bar{z}^*)$.
Finally, {\tt findpts\_eval} interpolates any given finite element function
$u$ at $\bx_j^*$ using the computational coordinates returned by {\tt findpts} and
a form similar to \eqref{eq_x}. The user is referred to Section 2.3 of
\cite{mittal2019nonconforming} and Section 3.2 of \cite{mittal2019highly} for more details.
For brevity, we will refer to interpolation of any scalar function
$u_B(\bx_B)$ at
arbitrary points ($\bx^*$) in physical space using \emph{findpts} as:
\begin{equation}
\label{eq_interpolation}
u(\bx^*) = \mathcal{I}(\bx^*, \bx_B, u_B(\bx_B)).
\end{equation}

Using \eqref{eq_interpolation}, we compute the integrated fitting error on each face as
\begin{equation}
\label{eq_int_fit_err_variation}
e_{f \in \mathcal{F}} = \sum_{q=1}^{N_q} w_q \big(\mathcal{I}(\bx_q, \bx_B, \sigma_B(\bx_B))\big)^2 d\bx_q,
\end{equation}
where $N_q$ is the number of quadrature points on face $f$ and $\bx_q$ are the physical space coordinates of the $q$th
quadrature point.
From an implementation perspective, we query \emph{findpts} to get the interpolated value at
all the quadrature points simultaneously, and use the interpolated values for integration as needed.

\begin{figure}[tb!]
\begin{center}
$\begin{array}{ccc}
\includegraphics[height=0.14\textwidth]{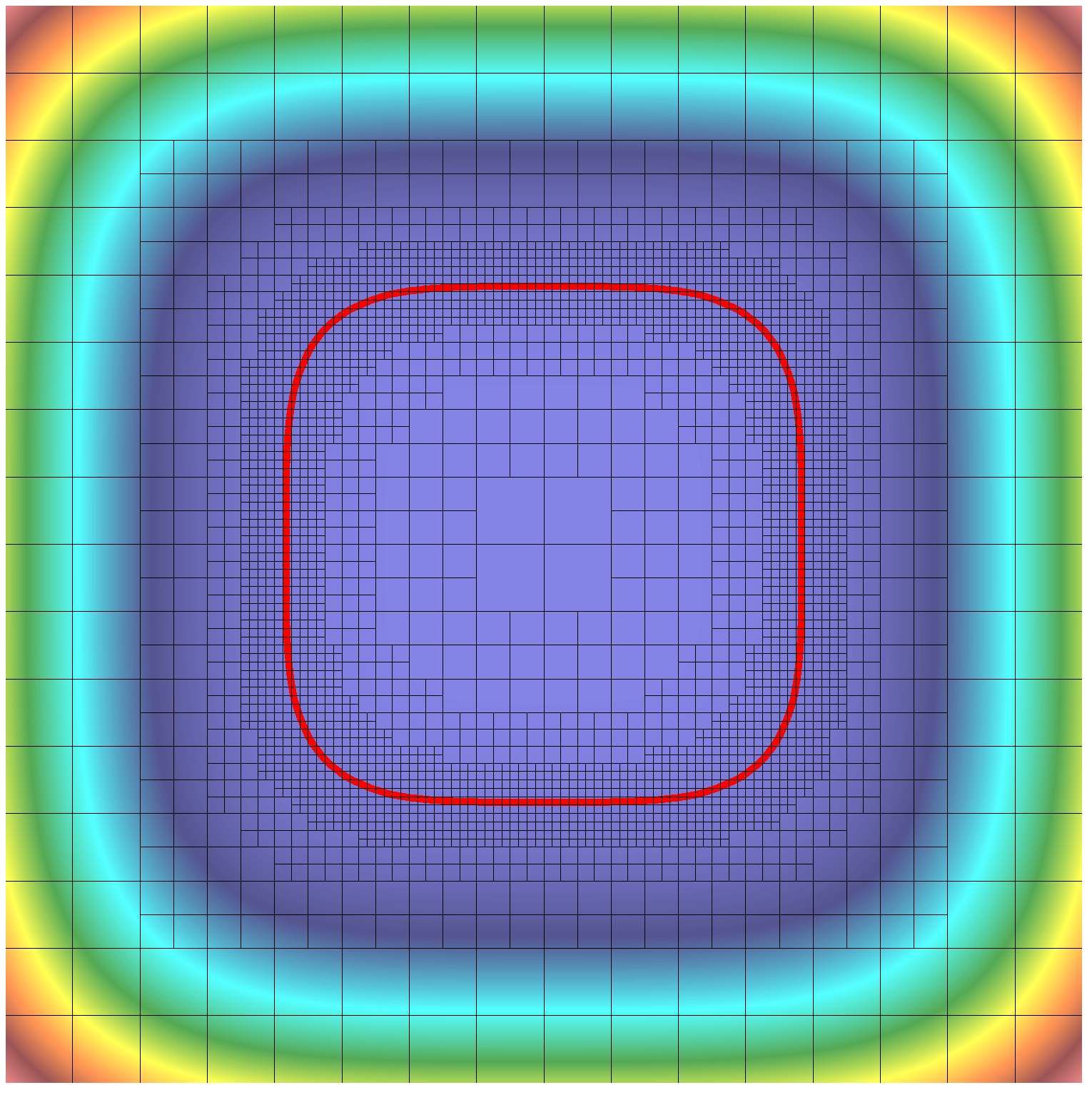} &
\includegraphics[height=0.14\textwidth]{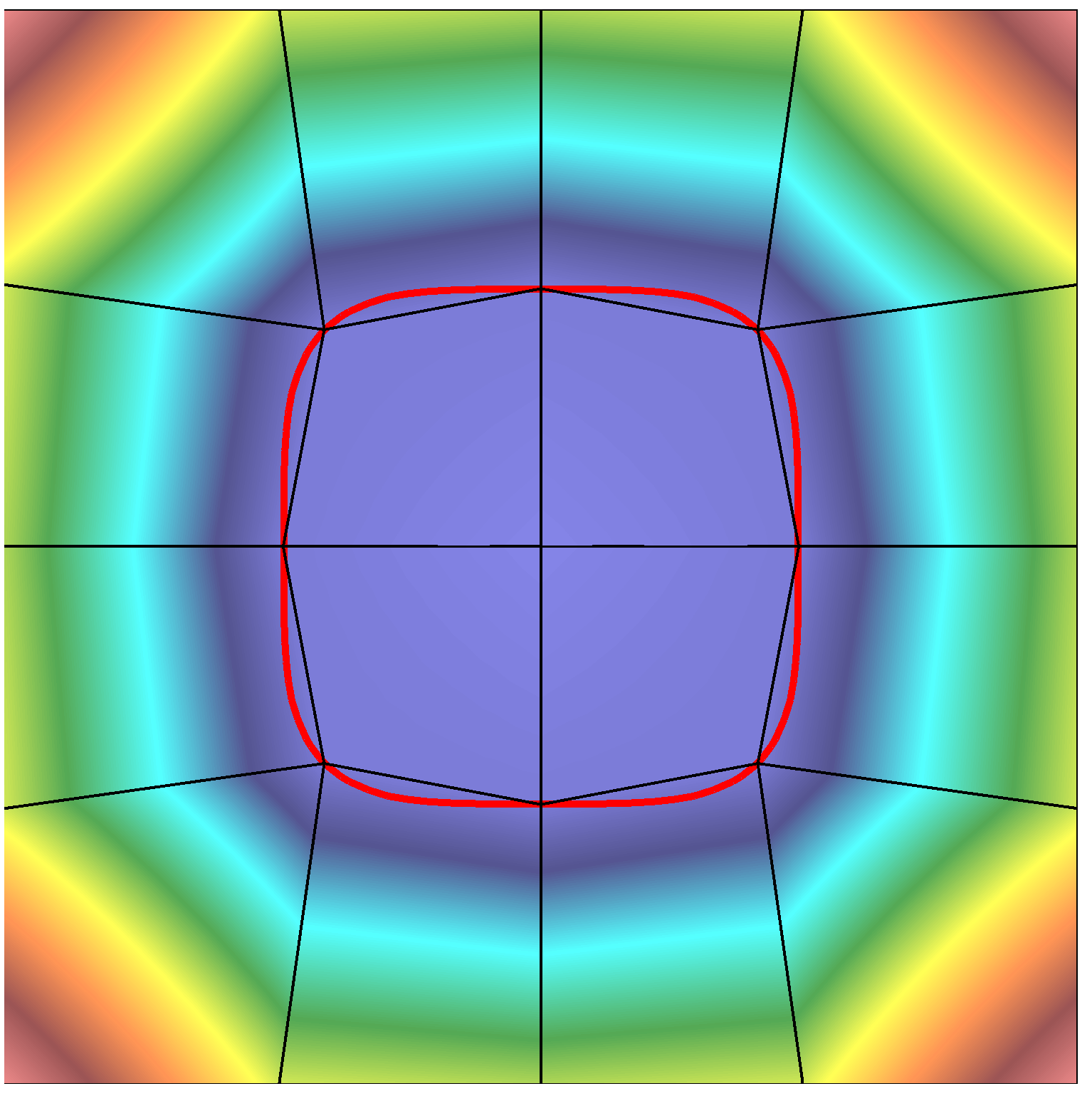} &
\includegraphics[height=0.14\textwidth]{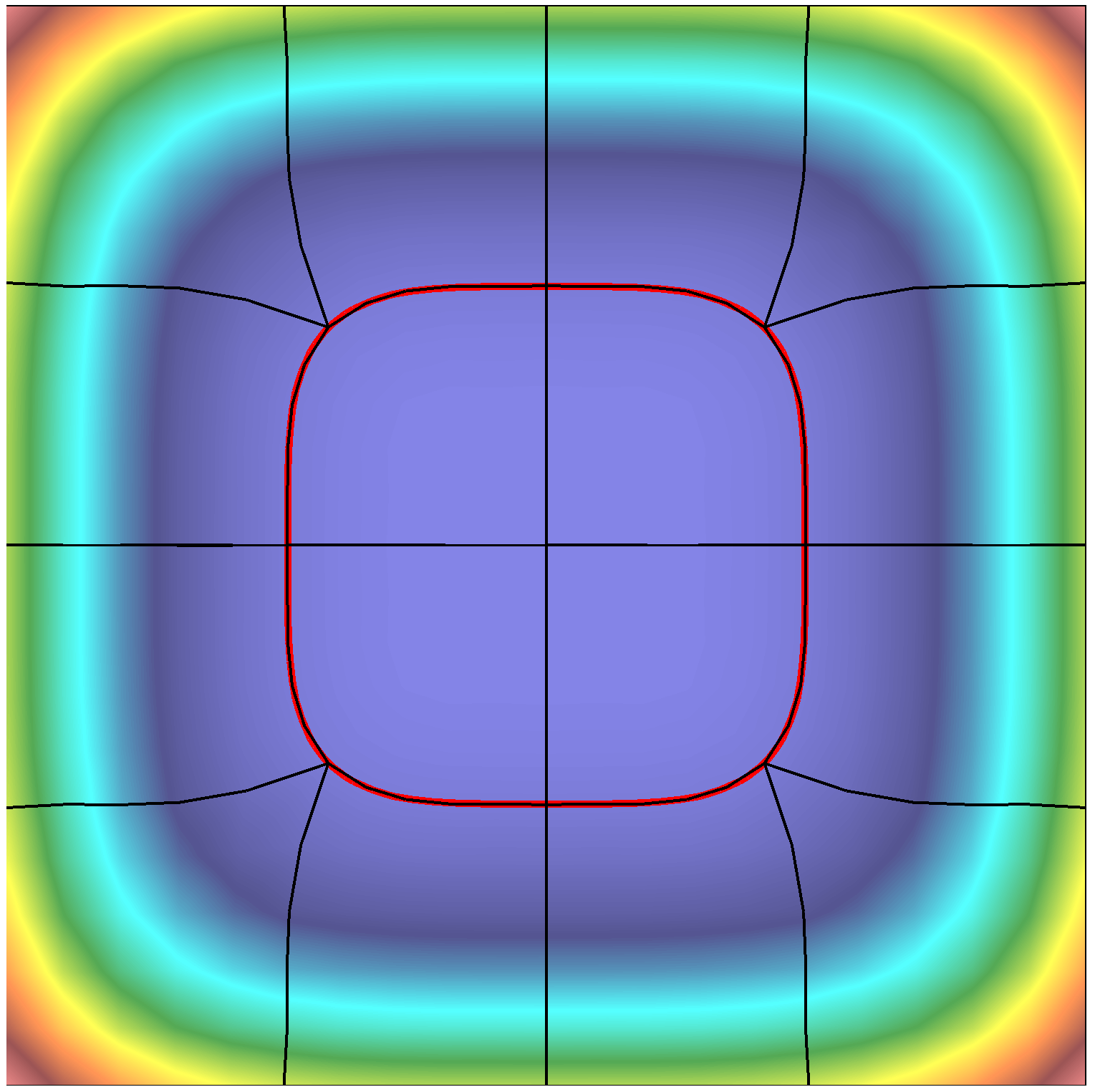} \\
\textrm{(a)} \sigma_B(\bx_B), p_{\sigma_B}=4 &
\textrm{(b) $\sigma(\bx), p=1$} &
\textrm{(c) $\sigma(\bx), p=3$}
\end{array}$
\end{center}
\vspace{-5mm}
\caption{Interpolating the level-set function from (a) adaptively refined
background mesh with $p_{\sigma_B}=4$ to (b) an optimized linear mesh and (c) an optimized
cubic mesh.}
\label{fig_sigma_remap}
\vspace{-3mm}
\end{figure}

Figure \ref{fig_sigma_remap} shows an example of high-order interpolation through \emph{findpts}.
A quartic representation of the squircle level-set function is interpolated at the nodal positions
of the optimized linear- and cubic-mesh. The level-set function is of the same order as the mesh,
i.e. $p=1$ in Figure \ref{fig_sigma_remap}(b) and $p=3$ in Figure \ref{fig_sigma_remap}(c).
As expected, the $p=3$ case resembles the source function more accurately
than the $p=1$ case.

\subsection{$p$-Refinement Criterion for Interface Elements}
\label{subsec_pref}

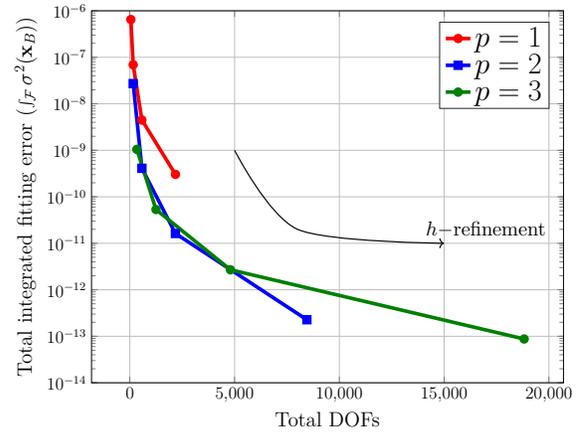
\begin{figure}[tb!]
    \centering
    \adjustbox{width=0.9\linewidth,valign=b}{\input{./figures/error_squircle_uniformp}}
    \vspace{-3mm}
    \caption{Comparison of total DOFs versus total integrated error for meshes with uniform polynomial orders ($p$). For each $p$
    considered, the coarse $4\times4$ mesh is uniformly refined up to 3 times, and aligned with the squircle
    level-set.}
    \label{fig_error_squircle_uniformp}
    \vspace{-3mm}
\end{figure}

To motivate our approach, we present  Figure \ref{fig_error_squircle_uniformp}
which shows how the total fitting error,
computed as the sum of integrated fitting error \eqref{eq_int_fit_err} over all the faces $f \in \mathcal{F}$
marked for fitting, varies with the number of degrees of freedom for different uniform-order optimized meshes.
For each $p$ considered, the coarse $4\times 4$ mesh shown in Figure \ref{fig_quad_squircle}(b) is also refined up to
3 times, each of which are indicated by different data points on the corresponding curve.
As we can see in Figure \ref{fig_quad_squircle}(d)-(f), with uniform mesh refinement,
we get a lot more elements away from the interface that
are essentially linear in shape, and using $p=3$ for these elements is increasing the computational
cost of the system without providing any gain in accuracy for the mesh alignment problem.
Also note that for a given number of DOFs, higher polynomial order results in higher accuracy, which
is why high-order meshes are preferred over uniformly refined low-order meshes.

The above observations are used to guide our approach for mesh $p$-refinement.
We typically start with the mesh at given polynomial
order, e.g., $p_{init} = 1$.
We then morph the mesh using the TMOP-based formulation \eqref{eq_F_full_sigma}.
Next, we compute the integrated fitting error $e_f$ \eqref{eq_int_fit_err} on each face $f$ marked for fitting,
and refine the elements adjacent to that face if the error is greater than a user prescribed threshold.
From a practical point of view, there are two choices for the refinement threshold (\erft). First is an absolute
threshold ($\gamma_1$) that is typically guided by the user's
knowledge of the level-set function. In this case, elements adjacent to a face are refined if
\begin{eqnarray}
\label{eq_ref_crit_one}
\textrm{Criterion 1: } e_f > e_{ref}=\gamma_1.
\end{eqnarray}
The second is a relative threshold (determined by $\gamma_2$) that depends on the maximum of
the integrated fitting error for all the
faces in the initial mesh ($e_{\mathcal{F},\infty}$).
In this case, adjacent elements are $p$-refined if
\begin{eqnarray}
\label{eq_ref_crit_two}
\textrm{Criterion 2: } e_f \geq \gamma_2 \cdot e_{\mathcal{F},\infty}.
\end{eqnarray}
Once the integrated fitting error has been computed for each face and the adjacent elements have been
marked for $p$-refinement, the polynomial order of these elements is increased by a user-prescribed
parameter (\delpr). Since not all the high-order nodes of this $p$-refined mesh
are located on the zero isocontour, the mesh is optimized again by minimizing \eqref{eq_F_full_sigma}.
This process of morphing followed by $p$-refinement is repeated until all the faces have their integrated fitting
error below the user-prescribed threshold or the elements have been elevated to maximum allowed polynomial order $p_{max}$.

In terms of increase in polynomial order $p$, there are two choices of interest. The first straightforward choice
is \delpr=1. A drawback of \delpr=1 is that it can require multiple Newton minimization steps for \eqref{eq_F_full_sigma}
(up to $p_{max}-p_{init}$ times) until the mesh aligns with the target isocontour with the desired accuracy (\erft).
Each of these Newton minimization steps has a non-trivial computational cost that increases with $p$.
The second choice is to use \delpr=$p_{max}-p_{init}$ such that all the interface elements to be refined
are elevated to the maximum polynomial order. The mesh can then be morphed to align with the level-set,
and no more Newton iterations would be needed afterwards.

Irrespective of what \delpr\ is, we first optimize the mesh at $p_{init}$, because the cost associated with it
can be significantly lower in comparison to higher polynomial orders if $p_{max}$ is much greater than $p_{init}$.
This choice is driven by the fact that the storage, assembly, and evaluation of FEM
operators scales as $\mathcal{O}(p^{2d})$, $\mathcal{O}(p^{3d})$, and $\mathcal{O}(p^{2d})$, respectively, for a
traditional full assembly-based approach \cite{MFEM2021}.
Furthermore, optimizing the mesh at a lower polynomial order, e.g., $p_{init}=1$, provides a good initial
condition for solving the high-order mesh optimization problem as demonstrated by Ruiz-Giron\'es et al. \cite{Roca2022}.

The approach outlined in this section is general in the sense that it gives the user flexibility on a case-by-case basis.
Our preliminary experiments show that a robust choice is to start with $p_{init}=1$, set \delpr=$p_{max}-p_{init}$, and simply set $e_{\tt ref}=\gamma_1=0$ to refine all elements at the interface once the linear mesh has been aligned with the level-set function.


\subsection{$p$-Derefinement Criterion}
\label{subsec_deref}

After we $p$-refine the mesh as described in the previous section,
the morphed mesh could have some interface elements at higher polynomial
order than needed, and these elements will unnecessarily increase the
computational cost of the actual simulation.
To ensure that interface elements are of minimum $p$ required to achieve the desired accuracy,
we take the optimized $p$-refined mesh and check for possible derefinements.
This is done by taking the finite element corresponding to each face $f$, and projecting it to
lower order spaces between the current polynomial order and $p_{init}$ sequentially,
and finding the lowest polynomial order that maintains the desired accuracy.
Unlike the refinement case, the TMOP objective is not
minimized again by solving the mesh alignment problem after derefinement,
i.e., there is no $r-$adaptivity step after derefinement.
Here, we consider three derefinement criteria.
Our first criterion is a threshold relative to the refinement threshold:
\begin{eqnarray}
\label{eq_deref_crit_one}
\textrm{Criterion 1: }  e_{f,\hat{p}} < e_{\tt deref}=\beta_1 e_{\tt ref},
\end{eqnarray}
where $e_{f,\hat{p}}$ represents the integrated error of the face $f$ evaluated at polynomial order
$p_{init}<\hat{p}<p_{max}$, and $\beta_1$ is a scalar that controls the desired accuracy relative to the
refinement threshold.
The second criterion is based on maximum allowed change in the integrated fitting error relative to
the current error at order $p$, i.e.,
\begin{eqnarray}
\label{eq_deref_crit_two}
\textrm{Criterion 2: }  e_{f,\hat{p}} < (1+\beta_2) e_{f,p}.
\end{eqnarray}
For example, $\beta_2$ can be 0.05 if the user wants to allow up to 5\% increase for an acceptable derefinement.
The third and final criterion is based on maximum decrease in the size $l$
of the edge/face $f$ relative to its current size:
\begin{eqnarray}
\label{eq_deref_crit_size}
\textrm{Criterion 3: }  l_{f,\hat{p}} > (1-\beta_3) l_{f,p}.
\end{eqnarray}
This criterion is motivated by the fact that a derefinement step can change the
shape quality of the element.
A small decrease of $l$ suggests that the shape quality around
the face remains mostly the same, which is the desired configuration.

Note that when interface elements are derefined, we must also make sure that none of these elements or their neighbors
become inverted. This is especially important in high-curvature regions.
Hence if a derefinement step leads to an inverted element, the step is rejected.

\begin{figure}[tb!]
\begin{center}
$\begin{array}{cc}
\includegraphics[height=0.2\textwidth]{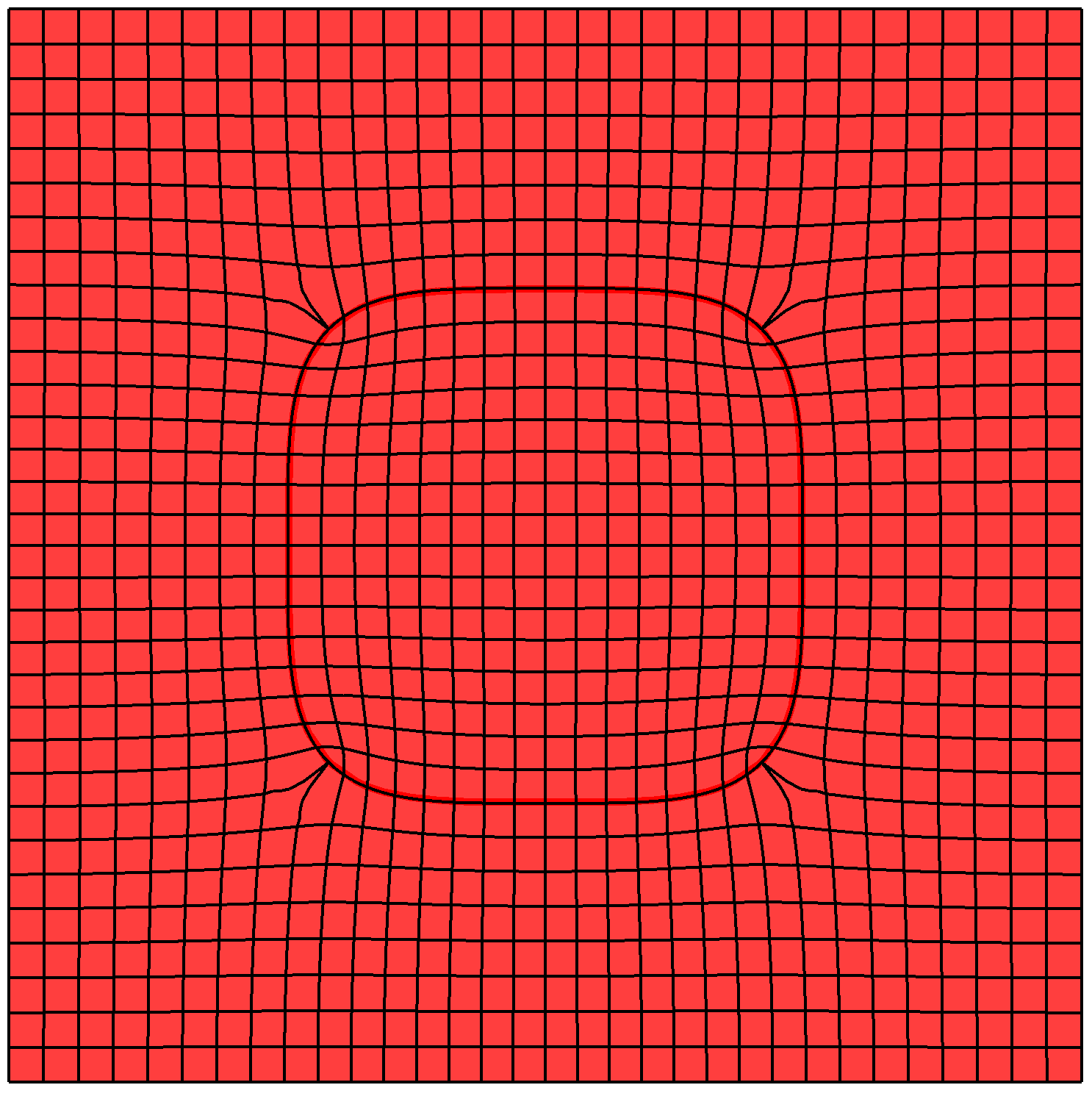} &
\includegraphics[height=0.2\textwidth]{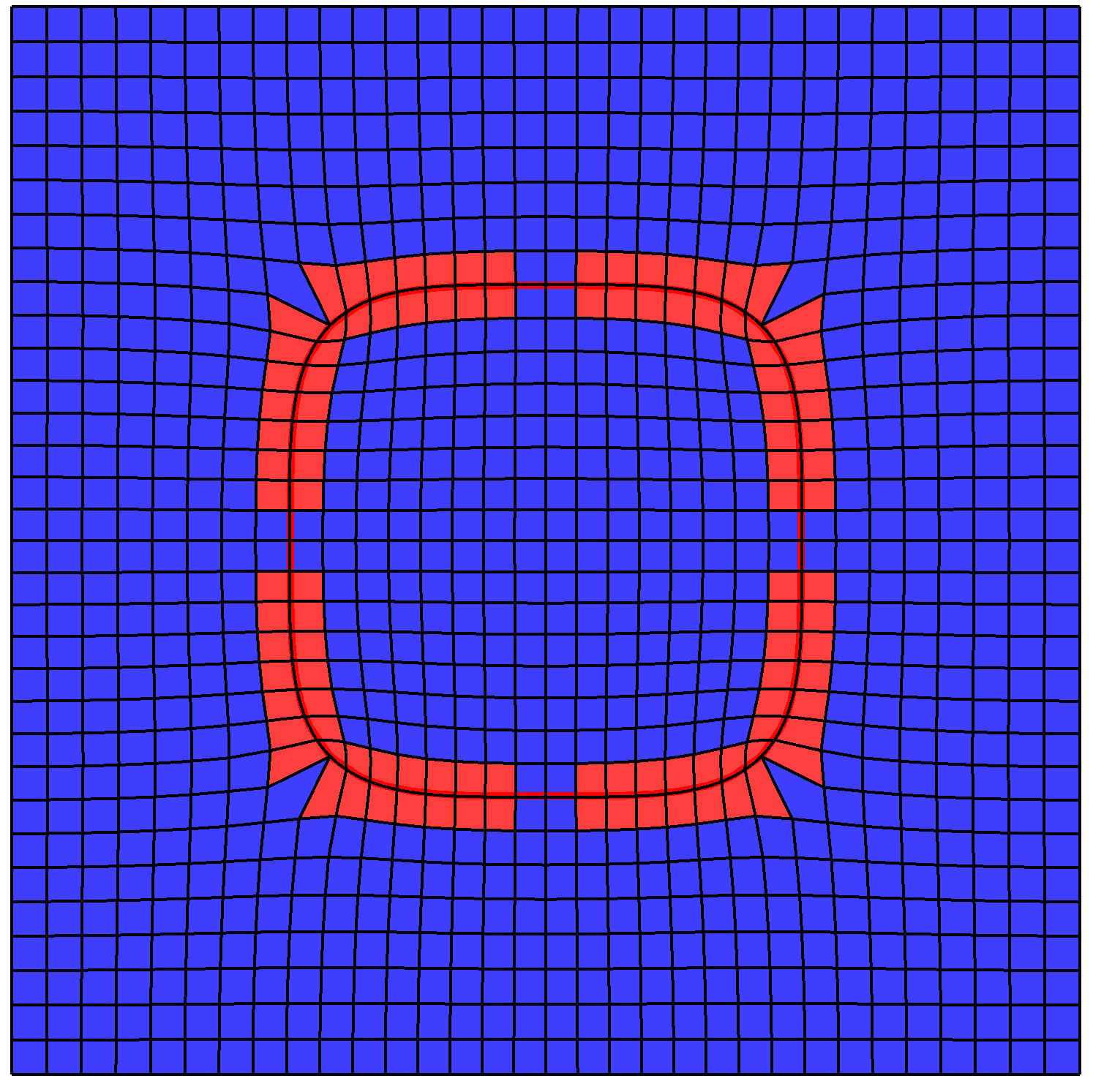} \\
\textrm{(a) Uniform $p=3$.} &
\textrm{(b) $p$-refined \textbf{A}.} \\
\includegraphics[height=0.2\textwidth]{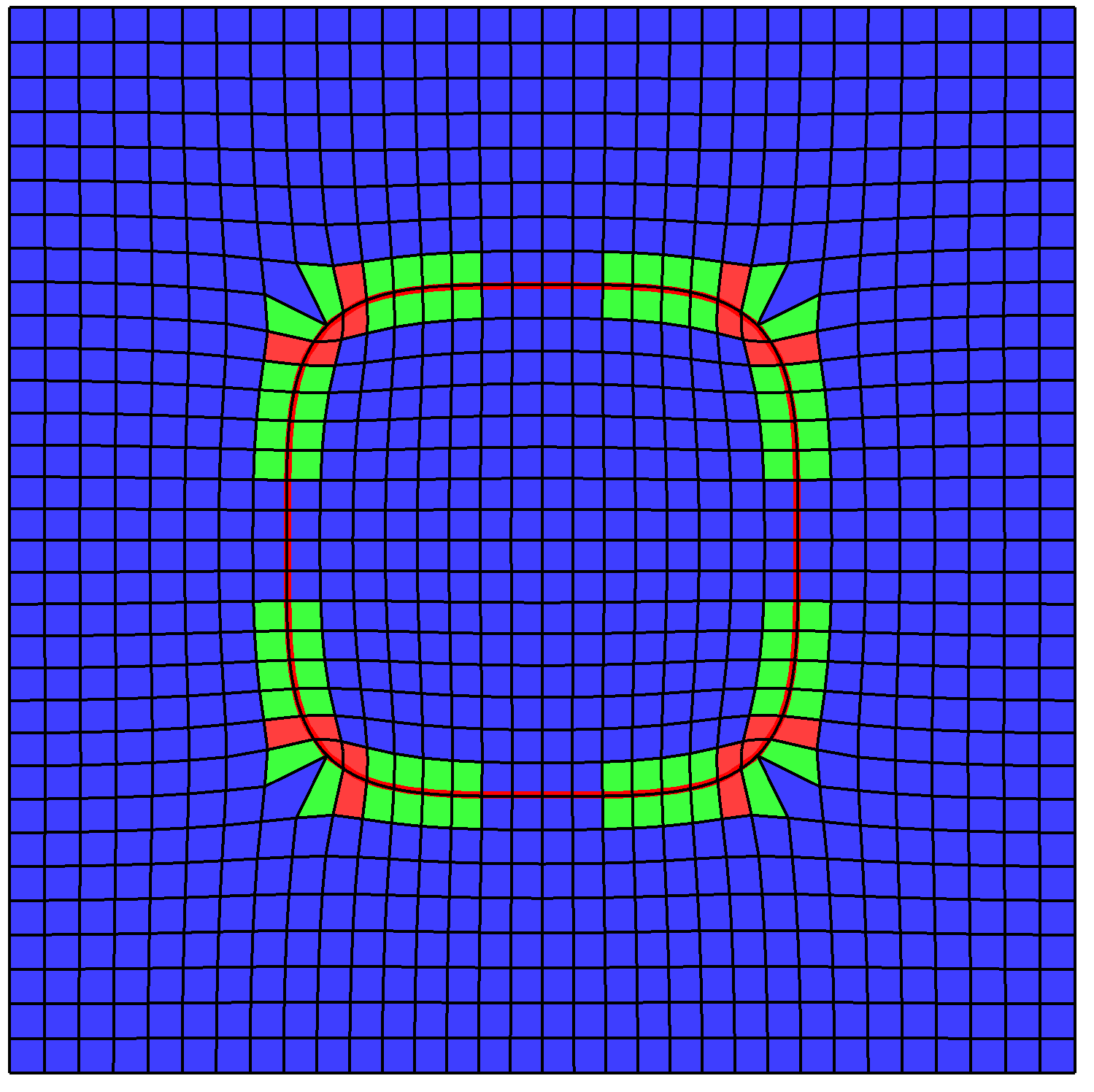} &
\includegraphics[height=0.2\textwidth]{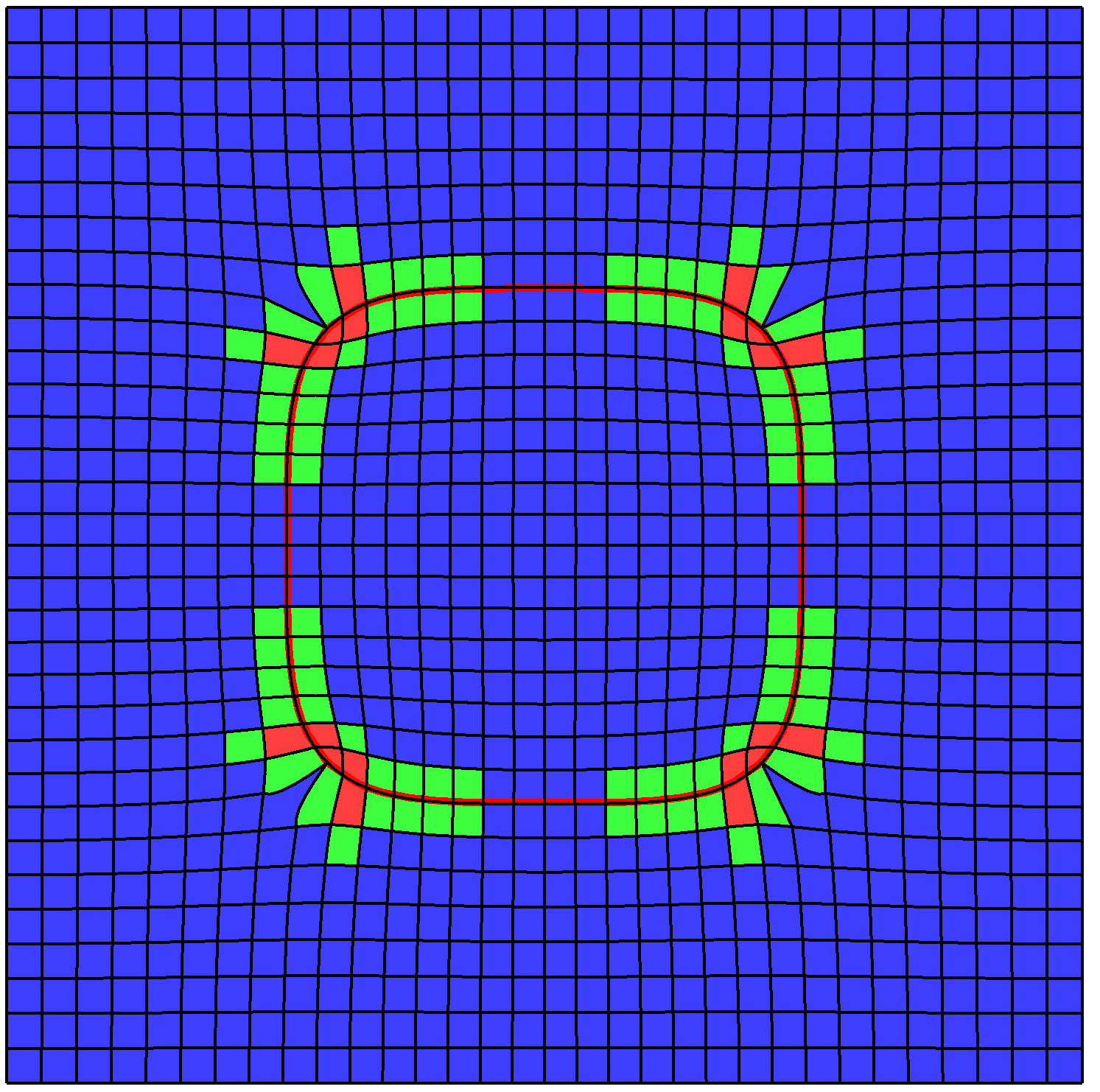} \\
\textrm{(c) $p$-refined \textbf{B}.} &
\textrm{(d) $p$-refined \textbf{C}.} \\
\multicolumn{2}{c}{\includegraphics[width=0.2\textwidth]{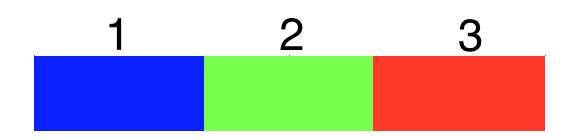}} \\
\multicolumn{2}{c}{\textrm{{\tt Polynomial order}}} \\
\multicolumn{2}{c}{\adjustbox{width=0.9\linewidth,valign=b}{\input{./figures/error_squircle_adaptivep}}} \\
\multicolumn{2}{c}{\textrm{(e) Error versus total DOFs for different approaches.}} \\
\end{array}$
\end{center}
\vspace{-5mm}
\caption{Comparison of uniform-order and $p$-refined meshes.
For uniform polynomial order, a coarse $4\times4$ mesh is uniformly $h$-refined up to 3 times, and aligned with the squircle level-set. For $p$-refined meshes, three different approaches are considered. We start with a linear mesh
and align it with the squircle level-set. For approach \textbf{A}, we then $p$-refine the mesh around the interface using
$\gamma_1=10^{-14}$ \eqref{eq_ref_crit_one}, and then re-align the mesh with the level-set. For approach \textbf{B}, we take
this mesh and derefine it using the size-based criterion \eqref{eq_deref_crit_size} with $\beta_3=10^{-5}$.
For approach \textbf{C}, we do the same as approach \textbf{B},
but also make sure that the maximum difference in order of adjacent elements is at-most 1 when the mesh is
$p$-refined/derefined.
}
\label{fig_quad_squircle_padaptive}
\vspace{-3mm}
\end{figure}

Figure \ref{fig_quad_squircle_padaptive} shows a comparison of the $p$-refined meshes with uniform-order
meshes for the squircle level-set. In this example, $p_{init}=1$, and we set $p_{max}=3$, \delpr=2,
$\gamma_1=10^{-14}$, and $\beta_3=10^{-5}$, In each case, we first optimize the linear mesh, then $p$-refine the mesh
using the absolute error-based threshold, and optimize the mesh again.
At this point,
we show the total DOFs vs integrated error for the aligned mesh as approach \textbf{A} in Figure \ref{fig_quad_squircle_padaptive}(b).
As we can see, approach \textbf{A} results in the mesh that has essentially the same accuracy as a uniform-order $p=3$ mesh
but at a much lower computational cost.
For approach \textbf{B} in Figure \ref{fig_quad_squircle_padaptive}(c),
we also derefine the mesh at the end using $\beta_3=10^{-5}$.
This slightly reduces the total DOFs and
accuracy of the alignment as expected. Notice that the optimized mesh has highest order elements of $p=3$ at the high-curvature
region, and $p=2$ and $p=1$ at regions of low and no curvature, respectively.
Figure \ref{fig_quad_squircle_padaptive}(b) also demonstrates an example of $p$-derefinement being rejected to prevent inverting an
element. At the high-curvature regions of the squircle, the elements on the outside of the level-set derefine from $p=3$ to 2,
but the element on the inside stays at $p=3$.

\subsection{Propagating $p$-refinement to Interior Elements}
\label{subsec_prop_pref}

As pointed out in \cite{karman2022mixed}, for cases like viscous flow simulations where elements near
the boundary have high aspect-ratio, mesh curving of only the boundary elements
can tangle the mesh.  To support mesh curving for such cases, we allow the user to
specify the maximum allowable difference (\delp) between the polynomial orders of neighboring elements.
This parameter can vary as one moves away from the boundary
or simply be held constant. For example, for thin boundary layer elements, a user may want to impose \delp=0 for
the first few layers adjacent to the boundary and \delp=1 away from the boundary. On the other hand,
for the case like the squircle, \delp\ can be simply set to $p_{max}$ so that only
the elements at the surface of interest are curved. For lack of a better choice, \delp\ can also be held constant at 1
so that the element polynomial orders go smoothly from $p_{max}$ to $p_{init}$ as we go away from the surface marked for fitting.
From a $p$-adaptivity perspective, this entails propagating the polynomial orders of the elements adjacent to $\mathcal{F}$ after
$p$-refinement, but before realigning with the level-set, and after $p$-derefinement.
We show the final optimized mesh for the squircle case with \delp=1 in Figure \ref{fig_quad_squircle_padaptive}(d) as
approach \textbf{C}.
As evident, using this approach does not increase the total number of DOFs significantly,
maintains the fitting accuracy, but can potentially circumvent deterioration in mesh quality.
Note from an implementation perspective, the polynomial orders are propagated in our framework through
edge-based connections in 2D and face-based connections in 3D.

\subsection{Summary of the $p$-adaptivity Algorithm}\label{subsec_algo}

Algorithm \ref{algo_fit} summarizes our $p$-adaptivity methodology for mesh alignment. There are three sets of
inputs to our method. The first set consists of the parameters that affect the mesh quality, i.e. the target matrix $W$,
the mesh quality metric $\mu(T)$, and the mesh $\mathcal{M}$ with nodal coordinates $\bx$ that is to be optimized.
The second set impacts the mesh alignment problem from a uniform-order
mesh's perspective. These inputs are the background mesh
$\mathcal{M}_B$ with nodal coordinates $\bx_B$ and the level-set function $\sigma_B(\bx_B)$ defined on it.
The first two sets together define the $r$-adaptivity method for mesh morphing to align with a level-set function \cite{barrera2023high}. The third and final set is the set of parameters that controls the $p$-adaptivity algorithm.
This consists of the maximum polynomial order allowed in the mesh $p_{max}$,
the change in element order upon $p$-refinement $\Delta p_{ref}$,
maximum allowed difference in polynomial orders of adjacent elements $\Delta p$,
and refinement- ($\gamma_1$ or $\gamma_2$) and derefinement-threshold
criterion ($\beta_1$ or $\beta_2$ or $\beta_3$).

\begin{algorithm2e}[h!]
\SetAlgoLined
\KwIn{$\mu$, $W$;\,\,\,
      $\mathcal{M}(\bx)$, $\mathcal{M}_B(\bx_B)$, $\sigma_B(\bx_B)$;\\
\,\,\, $p_{max}$, \delpr, $\Delta p$, ($e_{\tt ref}:\gamma_1 \vert \gamma_2$), ($e_{\tt deref}:\beta_1 \vert \beta_2 \vert \beta_3$)}
\KwOut{Variable order mesh aligned with level-set.}
Determine the set of faces $\mathcal{F}$ that have to be aligned with the level-set function.\\
Minimize \eqref{eq_F_full_sigma} to align the input mesh ($\bx$) to level-set function ($\sigma_B(\bx_B)$) while ensuring good mesh quality as prescribed by $\mu$ and $W$. See Algorithm 1 of \cite{barrera2023high}. \\
\While{true}{
    Compute $e_f \forall f \in \mathcal{F}$, and set polynomial order of faces with $e_f > e_{ref}$ to $\min(p_{max},p_f+\Delta p_{ref})$, Section \ref{subsec_remap} and \ref{subsec_pref}. \label{algo_pref_step} \\
    If no faces are $p$-refined, \textbf{go to } \ref{algo_end}.\\
    Propagate polynomial orders to interior elements, Section \ref{subsec_prop_pref}. \\
    Minimize \eqref{eq_F_full_sigma} for $r$-adaptivity.\\
    \If{$\Delta p_{ref}>1$} {
        Compute $e_f \forall f \in \mathcal{F}$ and decrease polynomial order of face to minimum order that meets derefinement criterion, Section \ref{subsec_deref}. \\
        Propagate polynomial orders to interior elements, Section \ref{subsec_prop_pref}. \\
    }
    If elements were refined up to $p_{max}$ at Step \ref{algo_pref_step}, \textbf{go to }\ref{algo_end}.
}
Return variable order mesh aligned with $\sigma_B(\bx_B)$.
\label{algo_end}
\caption{$rp$-adaptivity for Mesh Alignment}
\label{algo_fit}
\end{algorithm2e}


\section{Results}
\label{sec_results}

In this section, we demonstrate the impact of our $rp$-refinement approach using various numerical experiments
and show that it extends to different element types in 2D and 3D.
Our implementation uses the general finite element infrastructure
provided by the MFEM finite element library~\cite{MFEM2021,mfem-web}.

\begin{figure}[t!]
\begin{center}
$\begin{array}{cc}\hspace{-5mm}
\includegraphics[height=0.2\textwidth]{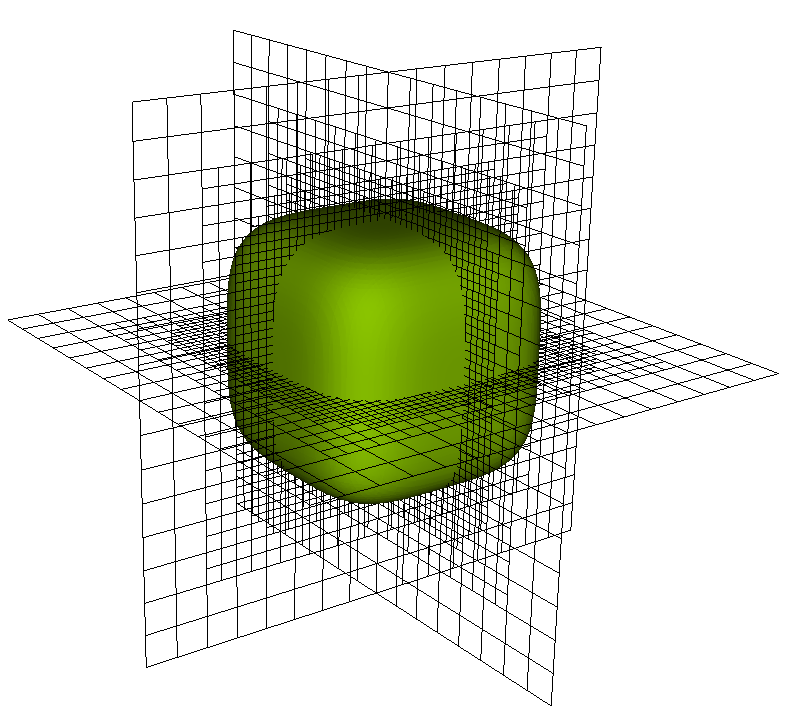} &
\includegraphics[height=0.2\textwidth]{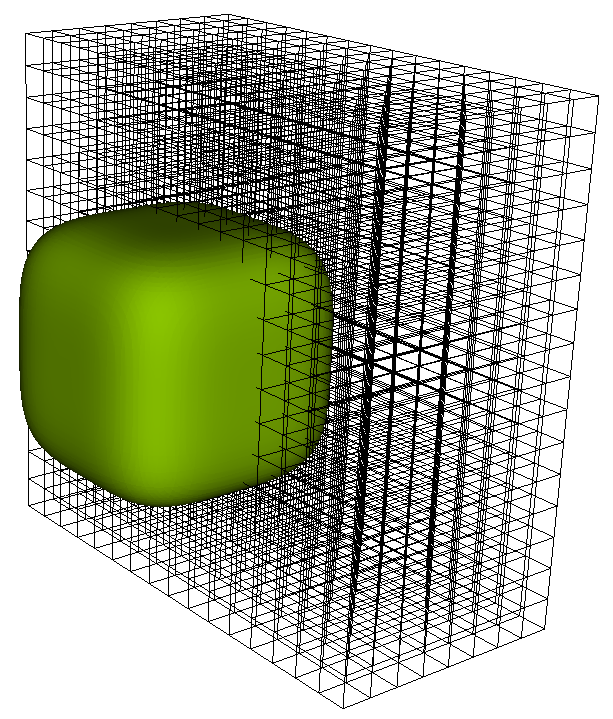}\\ \hspace{-5mm}
\vspace{-5mm} \\
\textrm{\textrm{(a)}}\,\mathcal{M}_B\,\,\textrm{and}\,\, \sigma(\bx_B)=0 & \textrm{(b)}\,\mathcal{M}\,\, \textrm{and}\,\, \sigma(\bx_B)=0 \\ \hspace{-5mm}
& N\approx 350k,\,\,e_{\mathcal{F}}=1.1\cdot 10^{-5}.  \\
\includegraphics[height=0.2\textwidth]{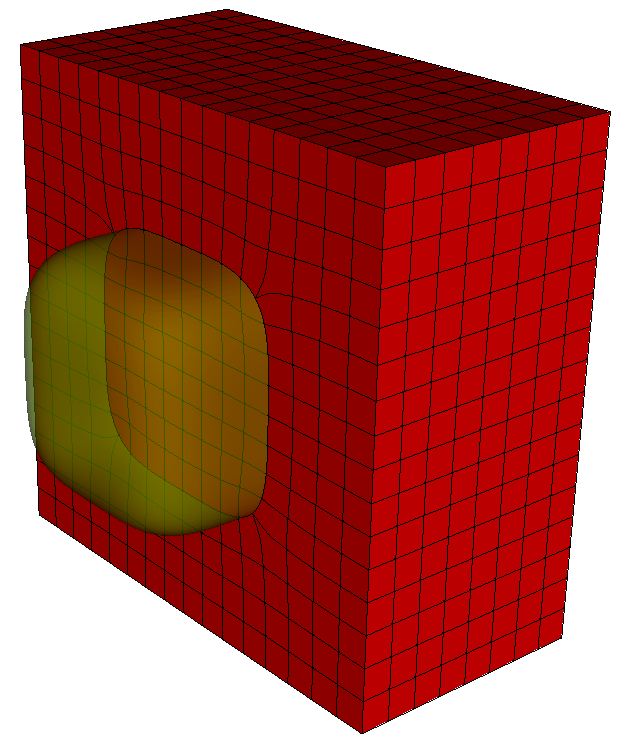} &
\includegraphics[height=0.2\textwidth]{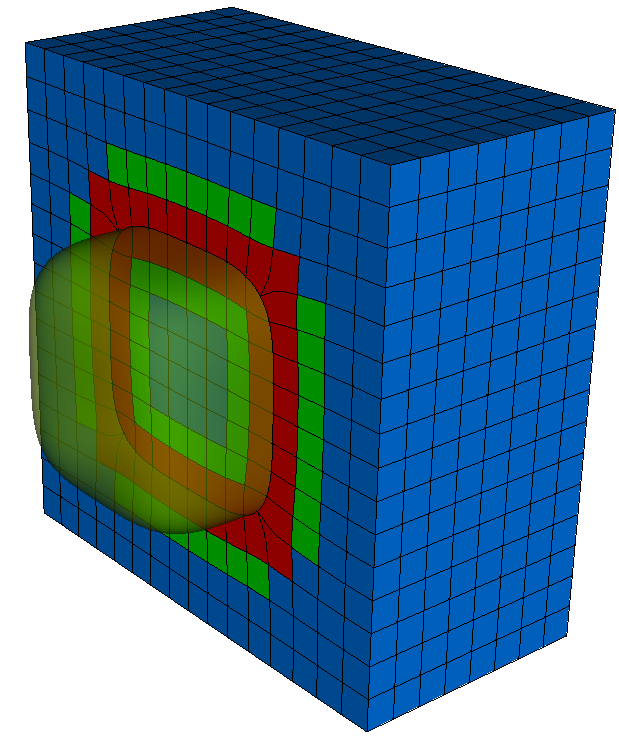} \\ \hspace{-5mm}
\vspace{-5mm} \\
\textrm{(c) Uniform}\,p=3. & \textrm{(d) Mixed-order mesh.} \\ \hspace{-5mm}
                           & \beta_3=0.0 \\ \hspace{-5mm}
N\approx 350k,\,\,e_{\mathcal{F}}=2.1\cdot 10^{-12}. & N\approx 102k,\,\,e_{\mathcal{F}}=2.1\cdot 10^{-12}. \\
\includegraphics[height=0.2\textwidth]{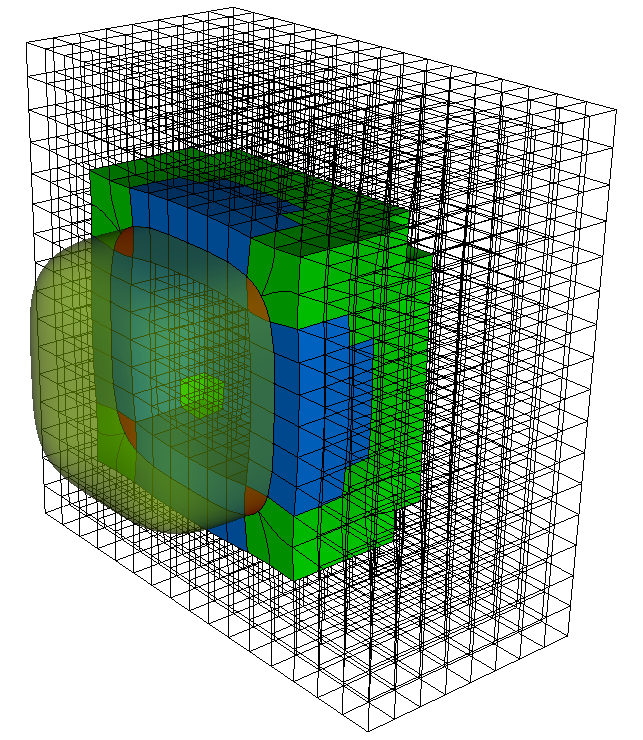} &
\includegraphics[height=0.2\textwidth]{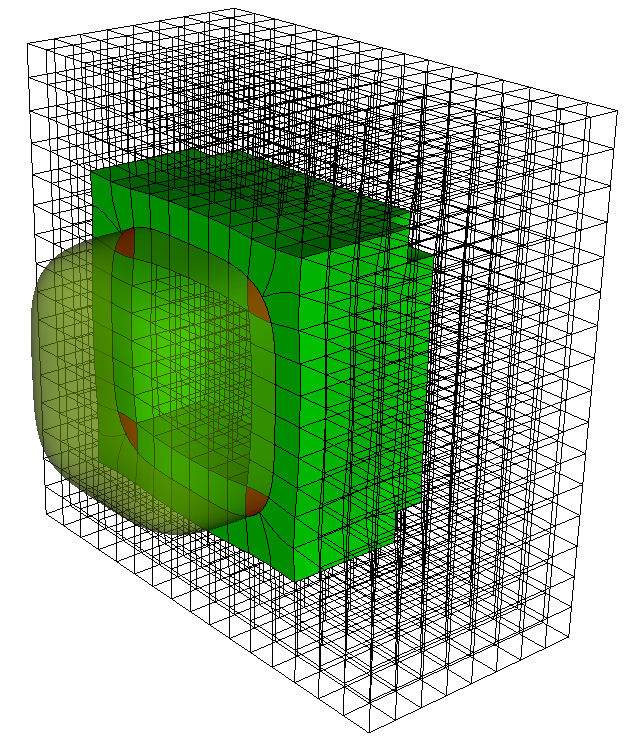} \\ \hspace{-5mm}
\vspace{-5mm} \\
\textrm{(e) Mixed-order mesh.} & \textrm{(f) Mixed-order mesh.} \\ \hspace{-5mm}
\beta_3=10^{-2} & \beta_3=10^{-3} \\
N\approx 35k,\,\,e_{\mathcal{F}}=1.7\cdot 10^{-9}. & N\approx 45k,\,\,e_{\mathcal{F}}=1.4\cdot 10^{-11}. \\
\multicolumn{2}{c}{\includegraphics[width=0.2\textwidth]{figures/legend_rgb_horizontal_1-3}} \\
\multicolumn{2}{c}{\textrm{{\tt Polynomial order}}} \\
\end{array}$
\end{center}
\vspace{-7mm}
\caption{Mixed order mesh curving for 3D. The zero isosurface of the level-set function
$\sigma(\bx_B)=0$ defined on the
background mesh ($\mathcal{M}_B$) is shown for each case, and only
the cut-view of $\mathcal{M}$ is shown for clarity.
For the $p$-refined meshes, we start with $p_{init}=1$ and use
$p_{max}=3$, $\Delta p=1$, $\gamma_1=0$. The final mixed-order meshes
are shown for $\beta_3=0$ (no $p$-derefinement), $10^{-2}$ and $10^{-3}$.
The number of DOFs ($N$) and fitting accuracy
$e_{\mathcal{F}}$ are indicated for the initial and optimized meshes.}
\label{fig_squircle3d_meshes}
\vspace{-3mm}
\end{figure}

\subsection{$p-$Refinement in 3D}
For our first example, we consider the 3D variant of the squircle level-set. The domain $\Omega \in [0, 1]^3$
is modeled using a $16\times16\times16$ uniform Cartesian mesh, and the target surface is prescribed through the level-set function:
\[
\sigma(\bx)=(x-0.5)^4 + (y-0.5)^4 + (z-0.5)^4 - 0.3^4,
\]
which is discretized on an adaptively refined background mesh
with $p_{\sigma_B}=4$. A slice-view of the background mesh is shown with
the zero isosurface of the level-set in Figure \ref{fig_squircle3d_meshes}(a), and a slice-view of the uniform linear
mesh to be optimized is shown in Figure \ref{fig_squircle3d_meshes}(b).

For mixed-order curving, we first take the linear uniform mesh, and morph it to align with the targt surface.
Then, we $p$-refine the mesh around the interface uniformly (i.e. $\gamma_1=0.0$) with $p_{max}=3$, $\Delta p=1$, and re-align it with the level-set.
Figure \ref{fig_squircle3d_meshes}(d) shows the $p$-refined mesh with $p=3$ around $\mathcal{F}$. Here, $\Delta p=1$,
and as a result the difference in polynomial order of face neighbors is 1.
Finally, we take this optimized mesh and derefine it using the size-based criterion \eqref{eq_deref_crit_size}.
Figure \ref{fig_squircle3d_meshes}(e) and (f) shows the mixed-order mesh with $\beta_3=10^{-2}$ and $10^{-3}$, respectively.

\begin{figure}[bt!]
\centering
\adjustbox{width=0.8\linewidth,valign=b}{\input{./figures/error_squircle3d_uniformp}}
\caption{Comparison of total DOFs versus total integrated error for uniform-order and $p$-refined meshes in 3D.
We consider $p$=1, 2, and 3 for uniform-order meshes. For $p$-refinement, we use $p_{init}=1$, $p_{max}=3$,
$\Delta p=1$, $\gamma_1=0$, and show the mesh for $\beta_3=0$, $\beta_3=10^{-2}$, and $10^{-3}$.}
\label{fig_error_squircle3d}
\vspace{-3mm}
\end{figure}
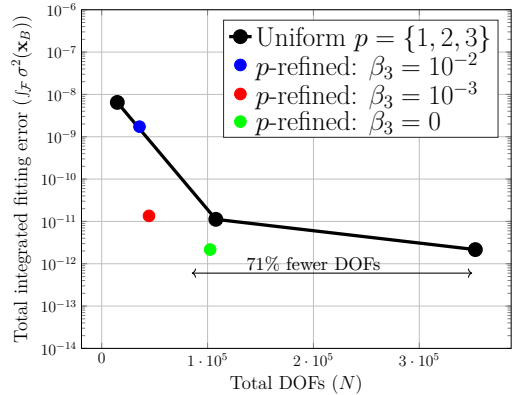

Figure \ref{fig_error_squircle3d} shows a comparison in the accuracy of alignment of the uniform-order meshes at $p$=1, 2, and 3,
with the $p$-refined meshes.
For the uniform-order meshes, increasing $p$ increases the computational cost significantly without much increase
in the geometric accuracy. This is similar to what we had observed in the 2D example.
For $p$-refined meshes, elevating the polynomial order of elements at the interface to $p_{max}=3$ with $\Delta p=1$
gives us the same geometric accuracy as a uniform $p=3$ mesh but at a 71\% lower computational
cost in terms of DOFs ($N=102435$ versus $352947$).
The $p$-refined mesh also has fewer DOFs than a
uniform-order quadratic mesh, and provides higher accuracy. Figure \ref{fig_error_squircle3d} also shows that derefining the mesh using the size-based criterion decreases the number of DOFs and the fitting accuracy as expected. The number of linear, quadratic, and cubic elements in each of the mixed-order meshes are summarized in Table \ref{tab_squircle3d_el_by_order}.

\begin{table}[h!]
\begin{center}
$\begin{array}{c}
\begin{tabular}{| c | c | c | c |}
\hline
 & p=1 & p=2 & p=3 \\
\hline
$\beta_3=0$ & 2784 & 536 & 776 \\
\hline
$\beta_3=10^{-3}$ & 3320 & 696 & 80 \\
\hline
$\beta_3=10^{-2}$ & 3704 & 312 & 80 \\
\hline
\end{tabular}
\end{array}$
\end{center}
\vspace{-4mm}
\caption{Number of linear, quadratic, and cubic elements for each of the mixed-order meshes aligned with the 3D level-set.}
\label{tab_squircle3d_el_by_order}
\end{table}

Recall that in our approach, we elevate polynomial order of elements
at the interface based on the integrated error on the faces that are
aligned to the level-set function (Section \ref{subsec_pref}).
Since the lowest-order entity constraints the solution, it is essential that
in 3D, elements having only edges on the interface are also $p$-refined when necessary (Section \ref{subsec_fem_pref}).
We handle such elements by setting their polynomial order as the
maximum of the polynomial orders of their neighbors that have a face on the interface.
This is essential for obtaining accurate alignment with the level-set in 3D.
Figure \ref{fig_squircle3d_meshes}(d) shows there are 4 such corner elements
in region of high curvature that have an edge on the interface,
which are elevated to $p=3$ due to their neighbors.

\subsection{Domain Prescribed by Geometric Primitives}
\begin{figure}[t!]
\begin{center}
$\begin{array}{cc}\hspace{-8mm}
\includegraphics[height=0.21\textwidth]{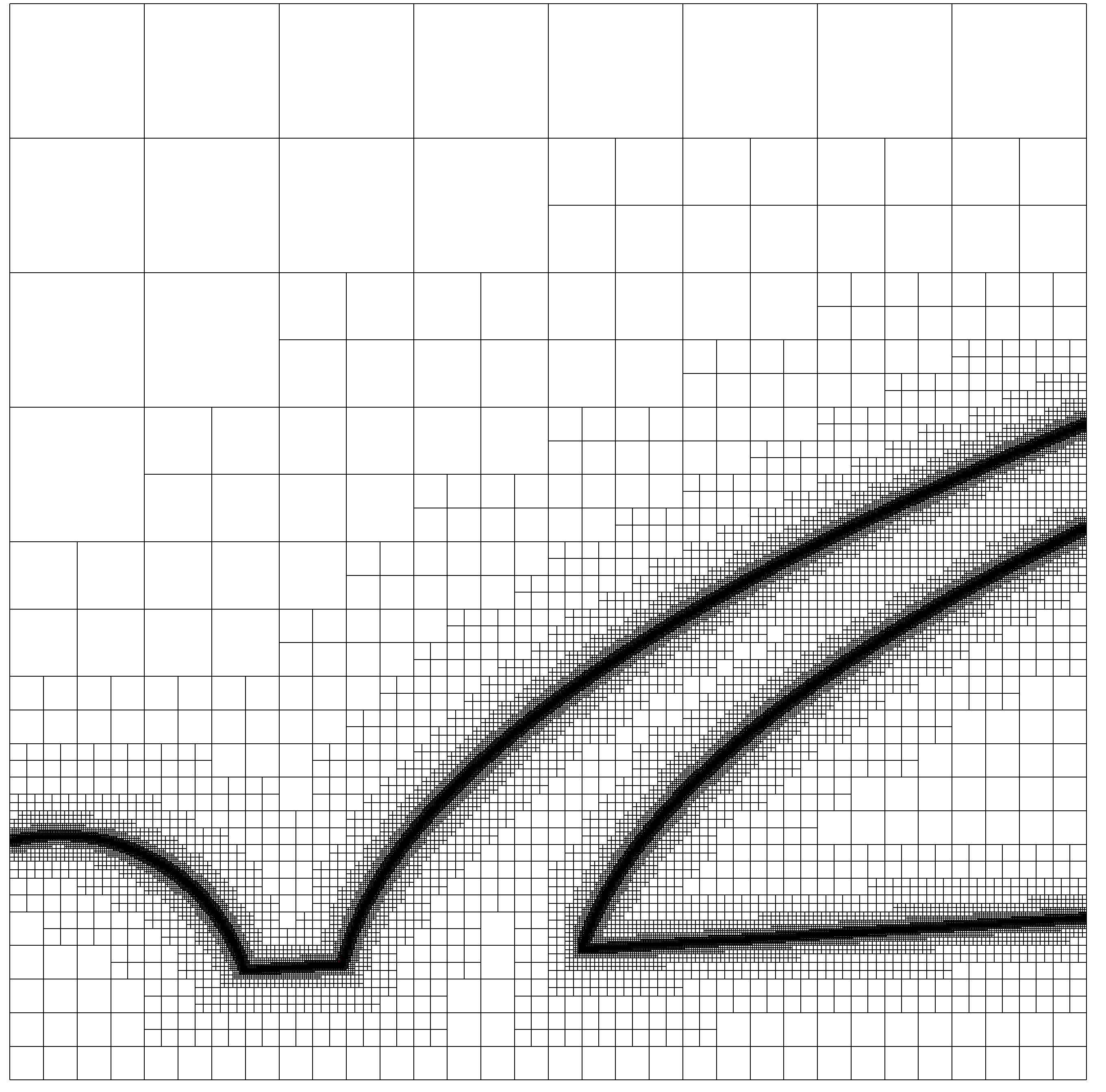} & \hspace{-4mm}
\includegraphics[height=0.21\textwidth]{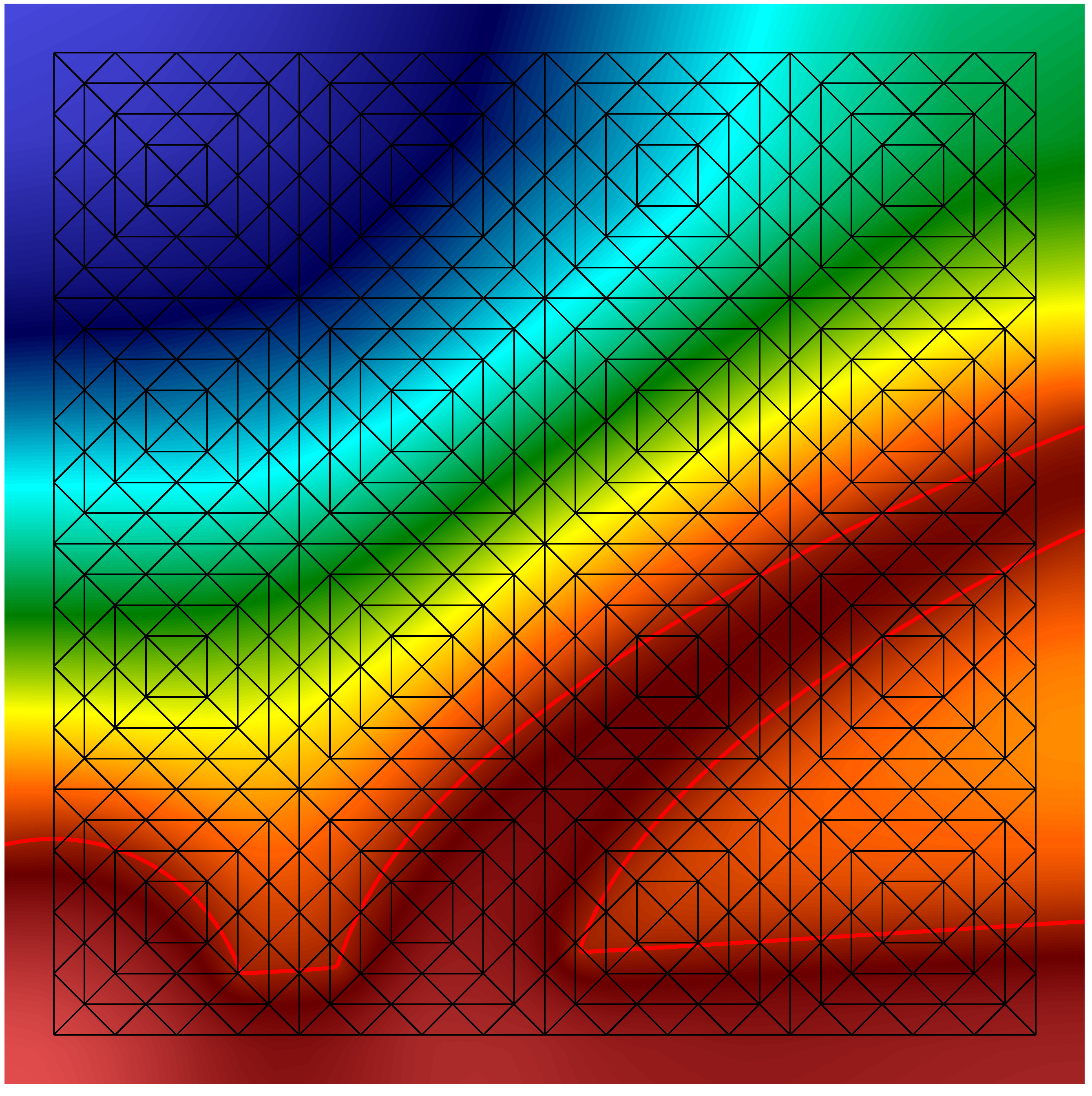} \\
\vspace{-5mm} \\
\hspace{-8mm}
\textrm{\textrm{(a)}}\,\mathcal{M}_B\,\, \textrm{and}\,\, \sigma_B(\bx_B)=0. &  \hspace{-4mm} \textrm{(b)}\,\mathcal{M}\,\, \textrm{and}\,\, \sigma_B(\bx_B); \Omega_B \supseteq \Omega. \\
\hspace{-8mm}
\includegraphics[height=0.21\textwidth]{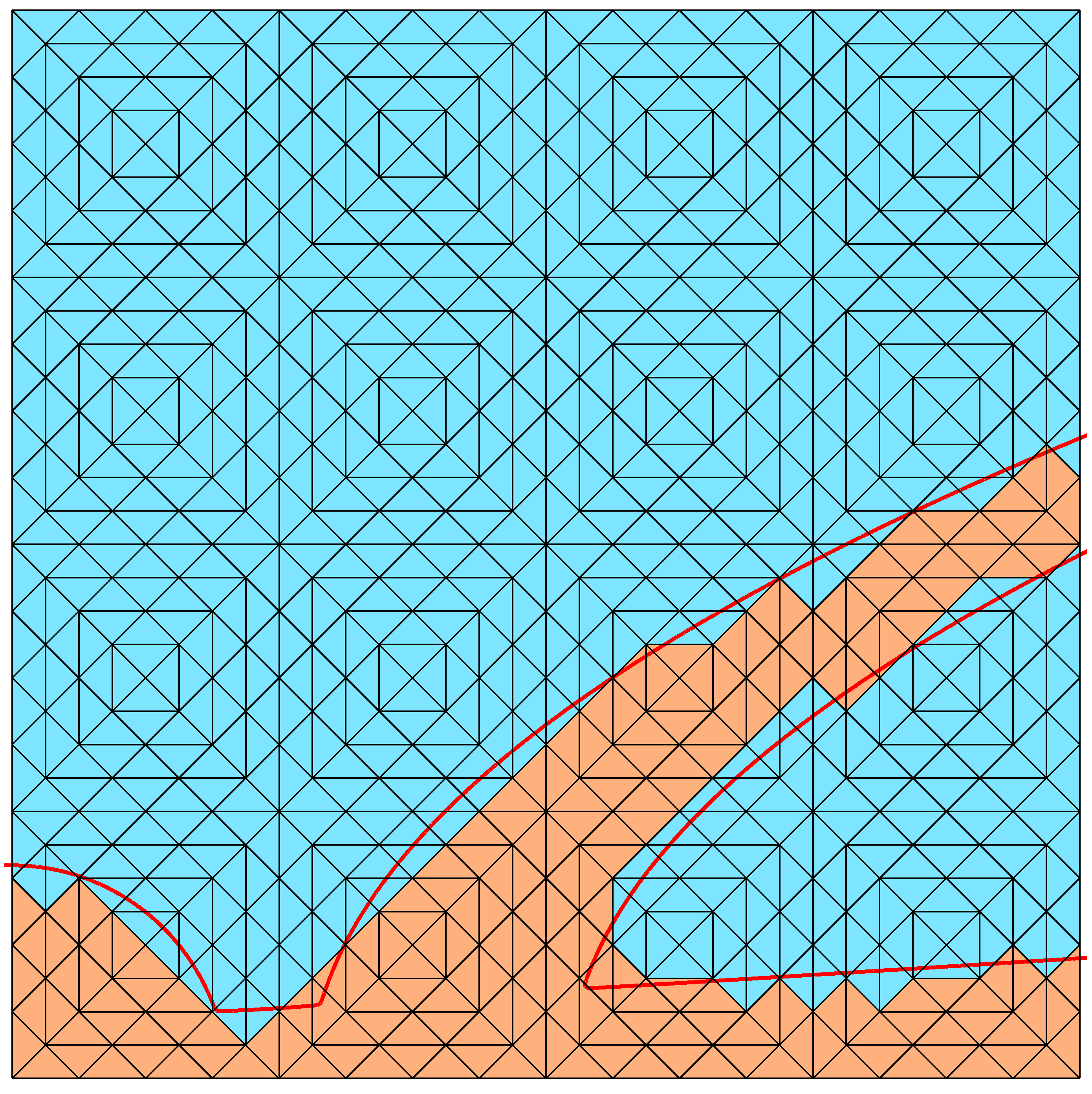} & \hspace{-4mm}
\includegraphics[height=0.21\textwidth]{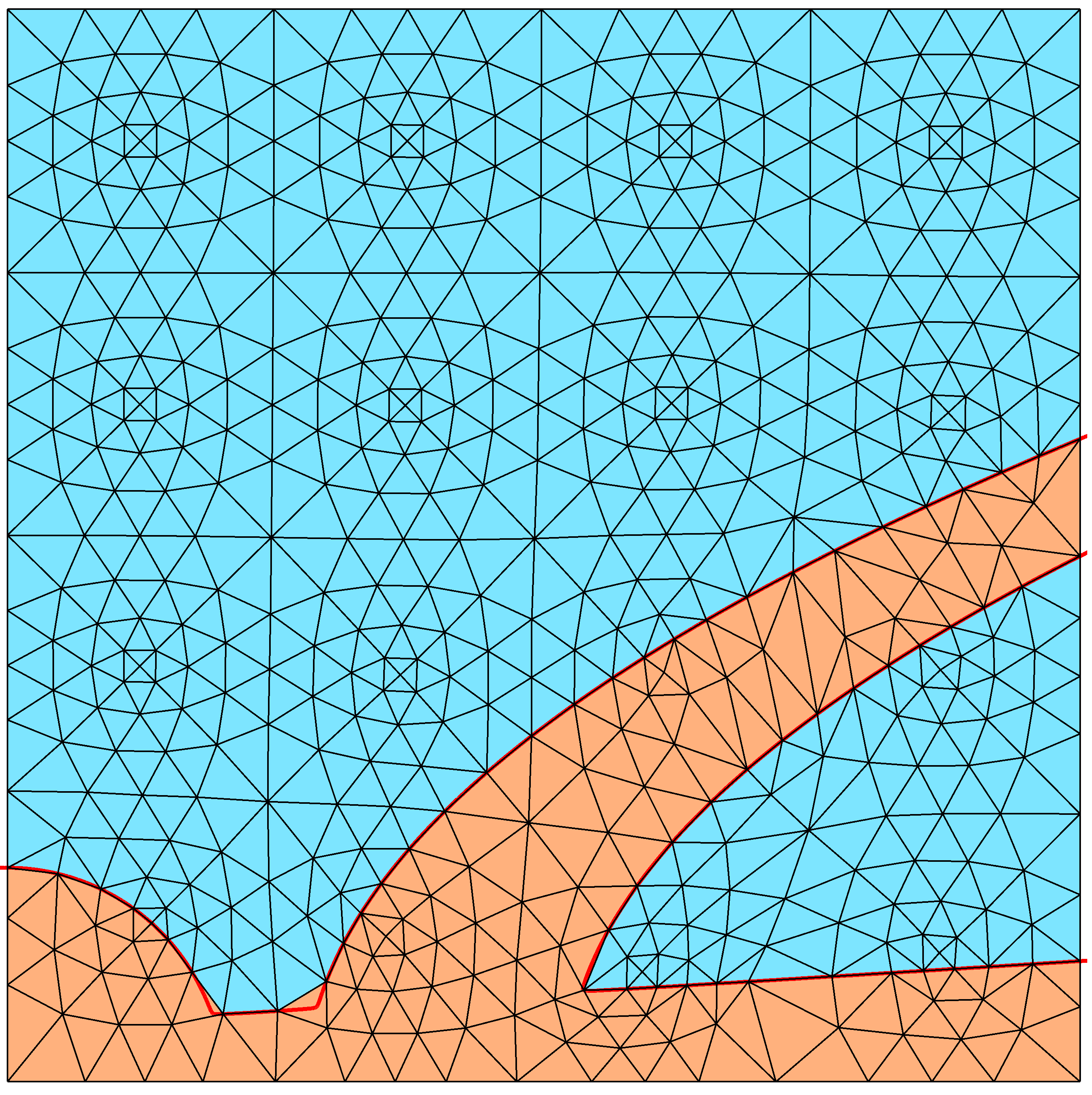} \\
\vspace{-5mm} \\
\hspace{-8mm}
\textrm{(c) Initial $\mathcal{M}$,}\,p=1. & \hspace{-4mm} \textrm{(d) Optimized $\mathcal{M}$,}\,p=1. \\ \hspace{-5mm}
N=1090, e_{\mathcal{F}}=2\cdot 10^{-5}. & N=1090, e_{\mathcal{F}}=2\cdot 10^{-6}. \\
\hspace{-8mm}
\includegraphics[height=0.21\textwidth]{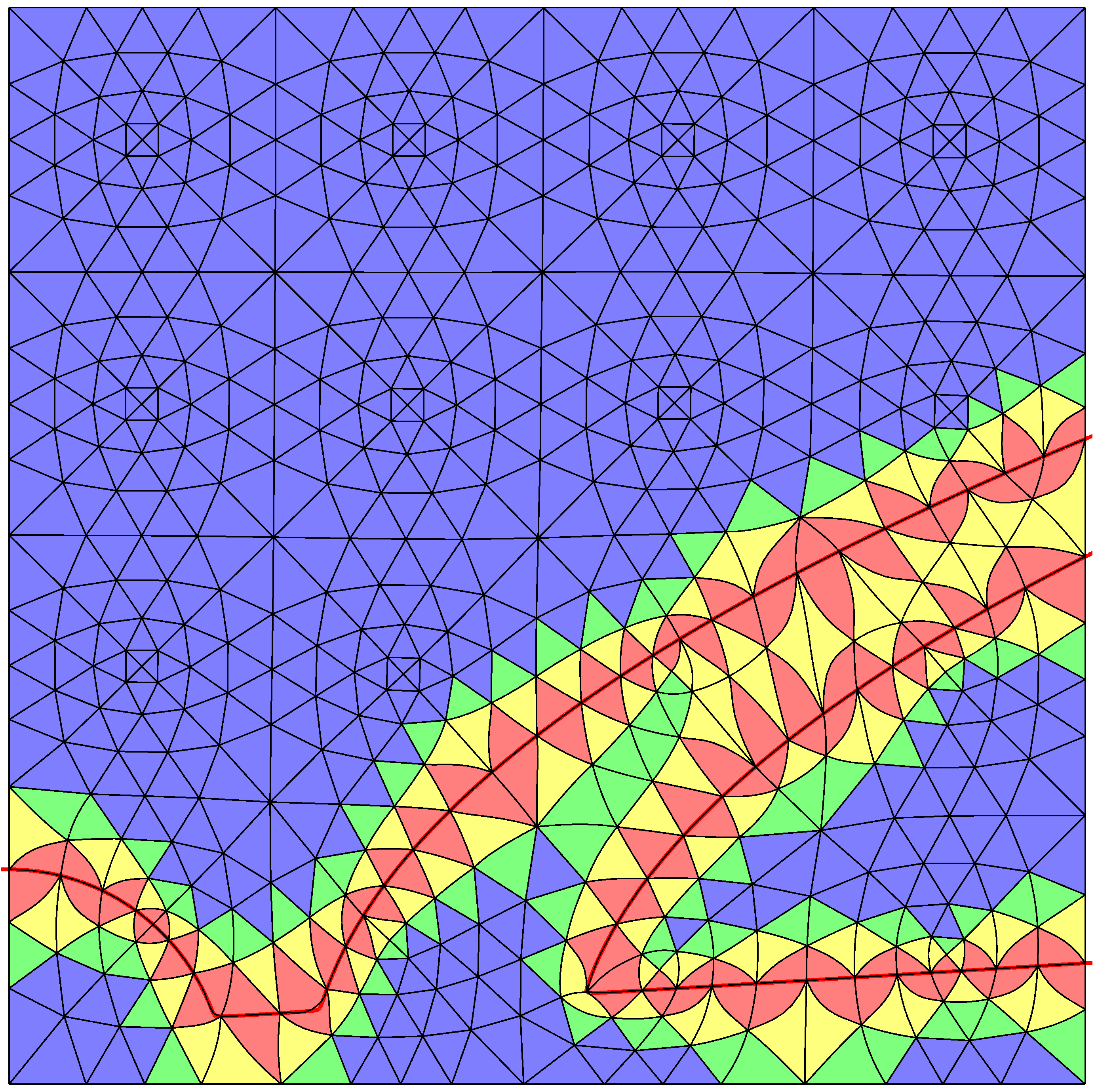} &
\hspace{-4mm} \includegraphics[height=0.21\textwidth]{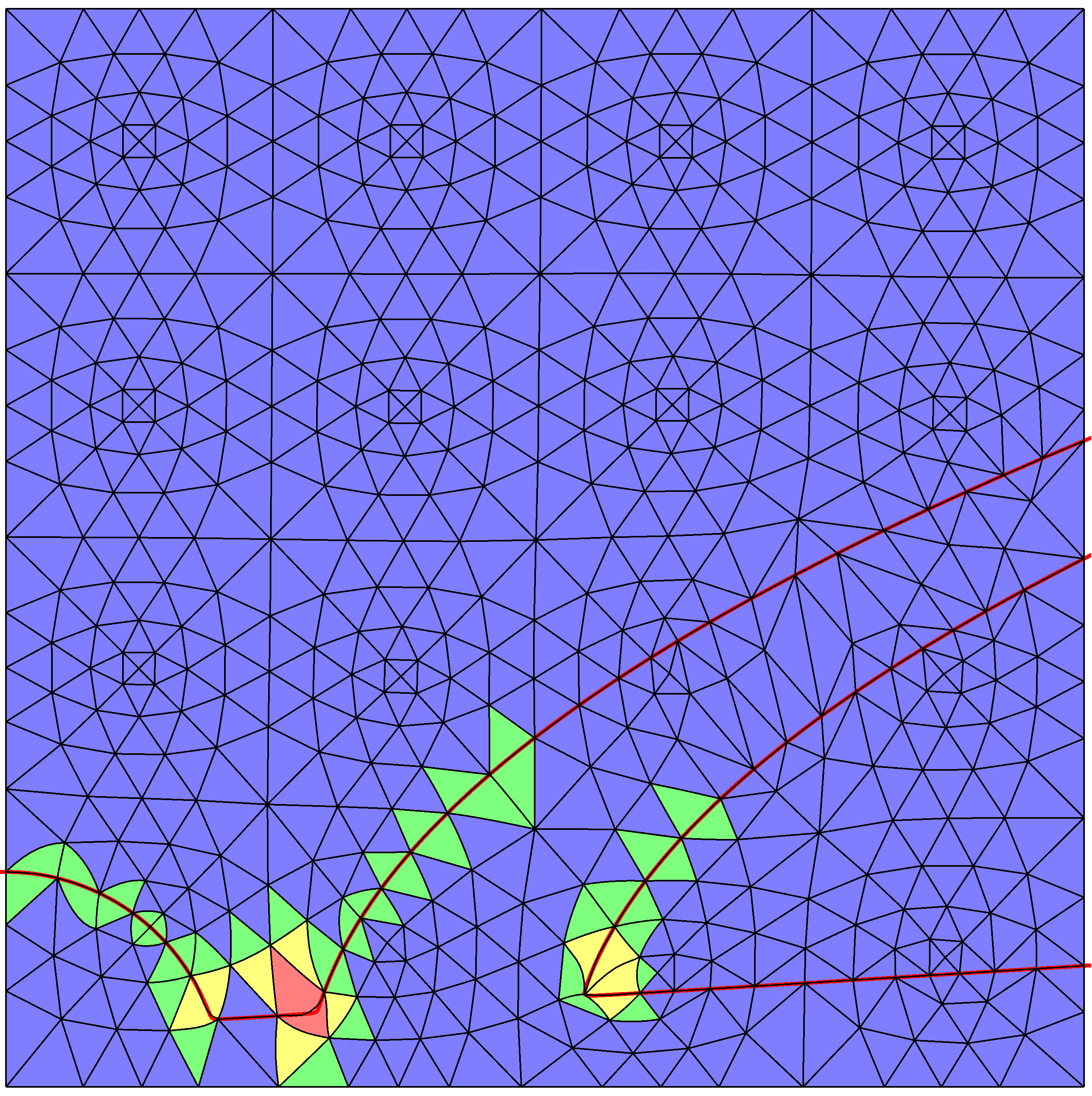} \\
\vspace{-5mm} \\
\hspace{-8mm}
\textrm{\textrm{(e) Optimized $\mathcal{M}$,}} & \hspace{-4mm}
\textrm{(f) Final mixed order $\mathcal{M}$.} \\
\textrm{$p=4$ around $\mathcal{F}$.} & \\
N=5086, e_{\mathcal{F}}=4\cdot 10^{-9}. & \hspace{-4mm} N=1472, e_{\mathcal{F}}=3\cdot 10^{-8}. \\
\multicolumn{2}{c}{\includegraphics[width=0.2\textwidth]{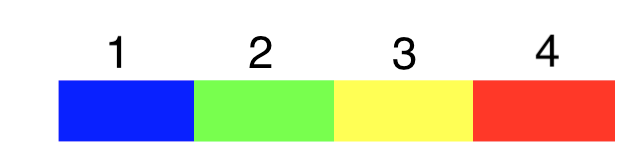}} \\
\multicolumn{2}{c}{\textrm{{\tt Polynomial order}}} \\
\end{array}$
\end{center}
\vspace{-7mm}
\caption{Mixed order mesh curving for a Fischer-Tropsch reactor like domain.
For the $p$-refined meshes, the maximum element order is $p_{max}=4$,
$\Delta p=1$, and the derefinement is done using size-based criterion
\eqref{eq_deref_crit_size} using $10^{-4}$.
The zero isosurface of the level-set function $\sigma_B(\bx_B)=0$ defined on the
background mesh $\mathcal{M}_B$ is shown for each case.
The number of DOFs ($N$) and fitting accuracy
$e_{\mathcal{F}}$ are indicated for the initial and optimized meshes.
}
\label{fig_reactor_padaptive}
\vspace{-3mm}
\end{figure}

Curvilinear domains are commonly prescribed as a combination of geometric primitives in
Constructive Solid Geometry (CSG) \cite{requicha1977constructive}. Figure \ref{fig_reactor_padaptive}
shows one such example adapted from \cite{barrera2023high} where the domain is prescribed
as an intersection of geometric primitive for a circle, parabola, and a trapezium.
Each geometric primitive is prescribed as a step function $\mathcal{G}(x_B)$
that is 1 inside one material and -1 inside the other material.
For such cases, we start with a coarse background mesh and
use adaptive mesh refinement around the zero level set of $\mathcal{G}(x_B)$.
When the zero isocontour interface intersects with the boundary of the domain,
as in the present case,  we configure the background mesh to completely
encompass the spatial domain of the morphed mesh $\mathcal{M}$,
and extend slightly beyond it, i.e. $\Omega_B \supseteq \Omega$.
This is essential for accurate computation of derivatives
near the boundary of $\mathcal{M}$.
Figure \ref{fig_reactor_padaptive}(a) shows the resulting
background mesh $\mathcal{M}_B$.
Next, we compute a discrete distance function $\sigma_B(\bx_B)$ using the
p-Laplacian solver of \cite{belyaev2015variational}, Section 7,
from the zero level set of $\mathcal{G}(x_B)$.
Figure \ref{fig_reactor_padaptive}(b) shows the mesh to be morphed $\mathcal{M}$,
the level-set function $\sigma_B(\bx_B)$, and its zero isocontour.

With the level-set function defined on $\mathcal{M}_B$, we assign the fictitious material indicators to elements
in $\mathcal{M}$ to determine element faces that will be aligned with $\sigma_B(\bx_B)$. The uniform
linear mesh with two materials is shown in Figure \ref{fig_reactor_padaptive}(c).
To minimize the TMOP problem, we use a shape metric with target transformation $W$ set to be that for an
equilateral triangle everywhere in the domain. The optimized linear mesh aligned with the zero isocontour of
$\sigma_B(\bx_B)$ is shown in Figure \ref{fig_reactor_padaptive}(d). Note that due to the prescribed mesh quality
metric and target, the elements away from the interface have optimized to as close to equilateral triangle
as possible. As expected, the optimized linear mesh aligns well with the zero isocontour in regions of low
curvature, but cannot capture the surface accurately in high curvature regions due to lack of degrees of freedom.

Next, using \erft=0., all the elements at the interface are refined to
$p_{max} = 4$ with $\Delta p = 1$.
This $p$-refined mesh is then morphed again to align with
the level-set function by minimizing the TMOP objective, as shown
in Figure \ref{fig_reactor_padaptive}(e).
Finally, we derefine the
elements based on the relative change in element size criterion ($\beta_3=10^{-4}$ in \eqref{eq_deref_crit_size}),
to obtain the mixed order mesh shown in Figure \ref{fig_reactor_padaptive}(f).

In terms of accuracy, the initial linear mesh has 1024 elements
with 1090 DOFs and a total integrated fitting error of $1.5\cdot 10^{-5}$.
This fitting error reduces to $1.5\cdot10^{-6}$ in the morphed linear mesh.
The $p$-refined mesh with quartic elements
at the interface has 5086 DOFs and $e_{\mathcal{F}}=3.8\cdot 10^{-9}$. Finally, the mixed order mesh after derefinement
has 967 linear elements, 45 quadratic elements, 10 cubic elements, and 2 quartic elements (Figure \ref{fig_reactor_padaptive}(f)), with a total of 1472 DOFs and a fitting accuracy of $e_{\mathcal{F}}=2.6\cdot 10^{-8}$.
In contrast, the total number of DOFs would have been 16642 had a uniform fourth order mesh been used. Thus, this final mixed order mesh with  has 91\% fewer DOFs in comparison to a uniform-order mesh.
Note that in this example, the primary
source of geometrical error are the \emph{sharp} corners in the surface of interest. Ideally, such regions
should be handled with higher mesh resolution, and we will explore augmenting our framework with $h$-refinement
in future work.


\subsection{Coarse Mixed-Order Mesh Starting From a Dense Low-Order Mesh}
While high-order meshes have become popular in recent times, a lot of
classical mesh generation tools still support only low-order meshes.
In this example, we show use of our framework to
obtain a coarse mixed-order mesh starting from a dense low-order mesh.

\begin{figure}[tb!]
\begin{center}
$\begin{array}{cc}\hspace{-5mm}
\includegraphics[width=0.23\textwidth]{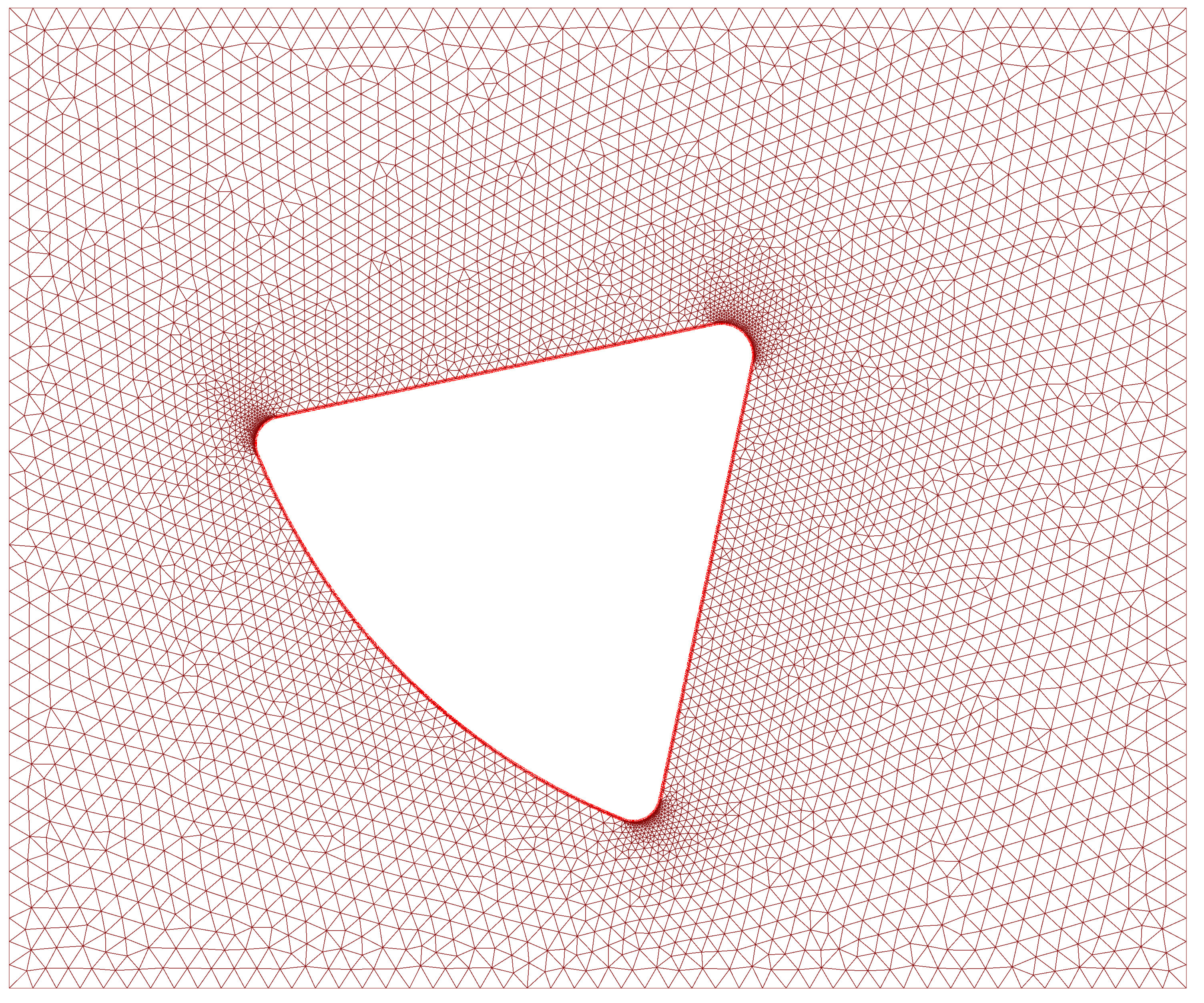} &
\includegraphics[width=0.23\textwidth]{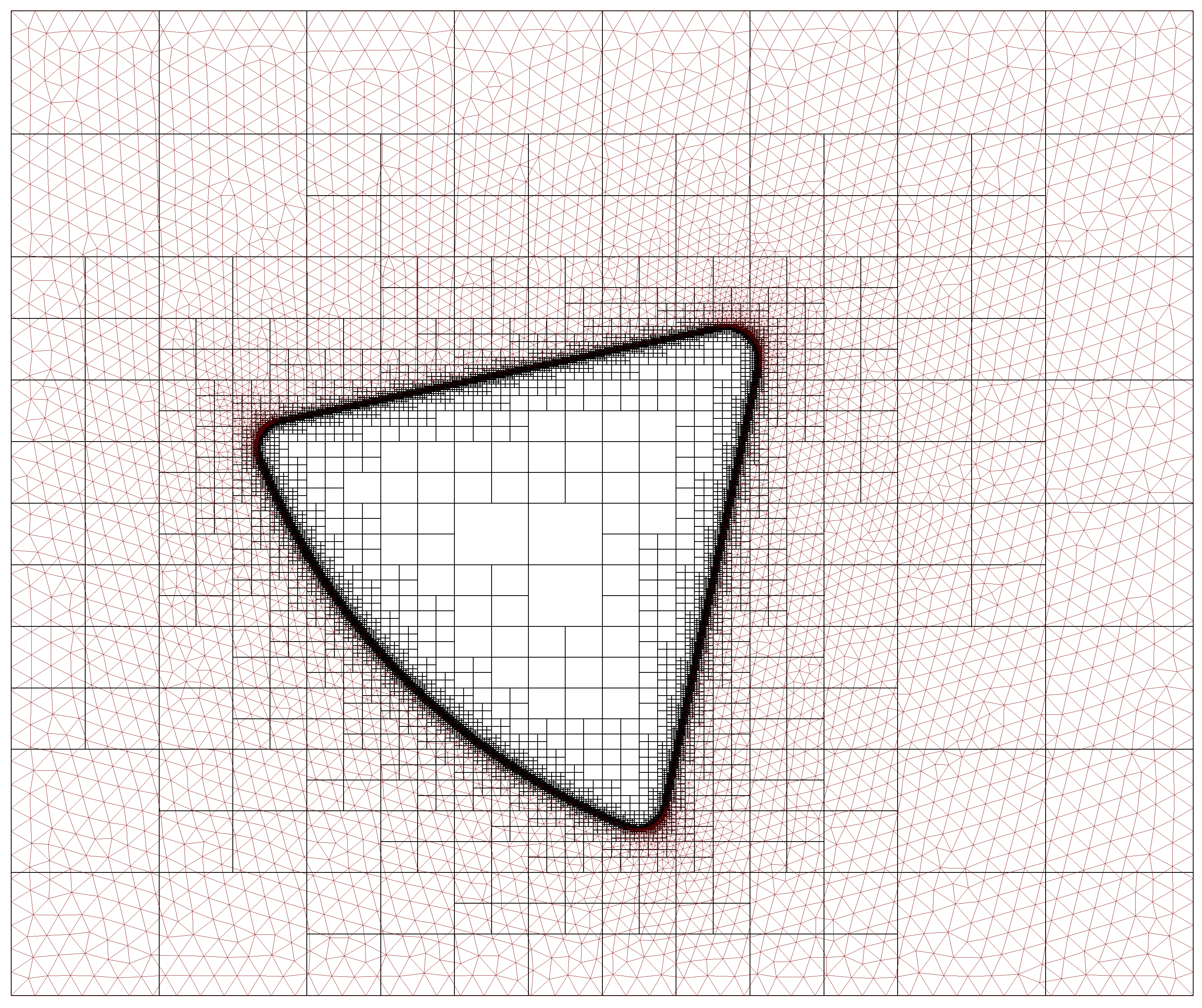} \\ \hspace{-7mm}
\vspace{-5mm} \\
\textrm{\textrm{(a)}}\,\mathcal{M}_{\tt lin} & \textrm{(b)}\,\mathcal{M}_{\tt lin}\,\, \textrm{and}\,\, \mathcal{M}_B \\ \hspace{-5mm}
\includegraphics[width=0.23\textwidth]{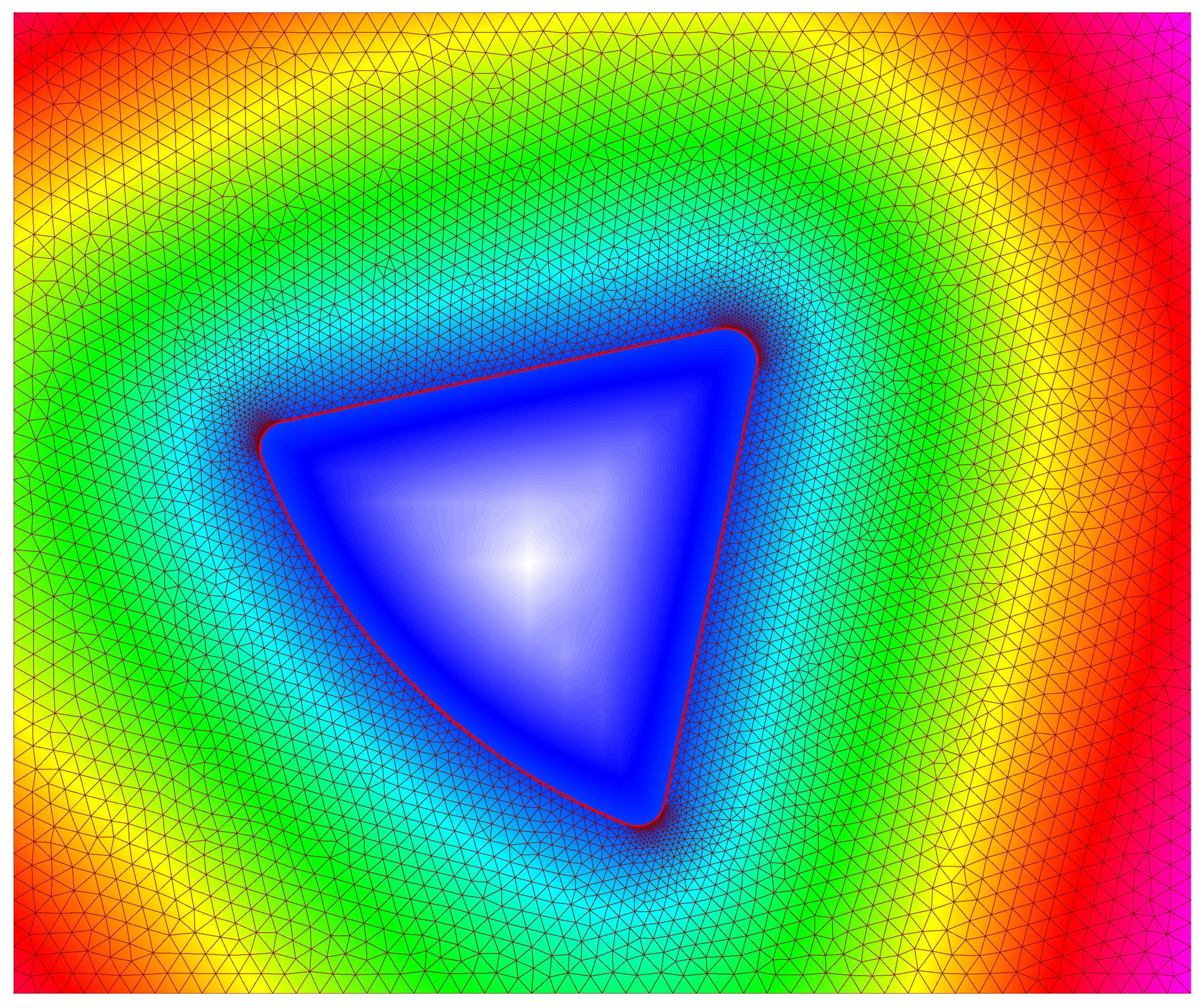} &
\includegraphics[width=0.23\textwidth]{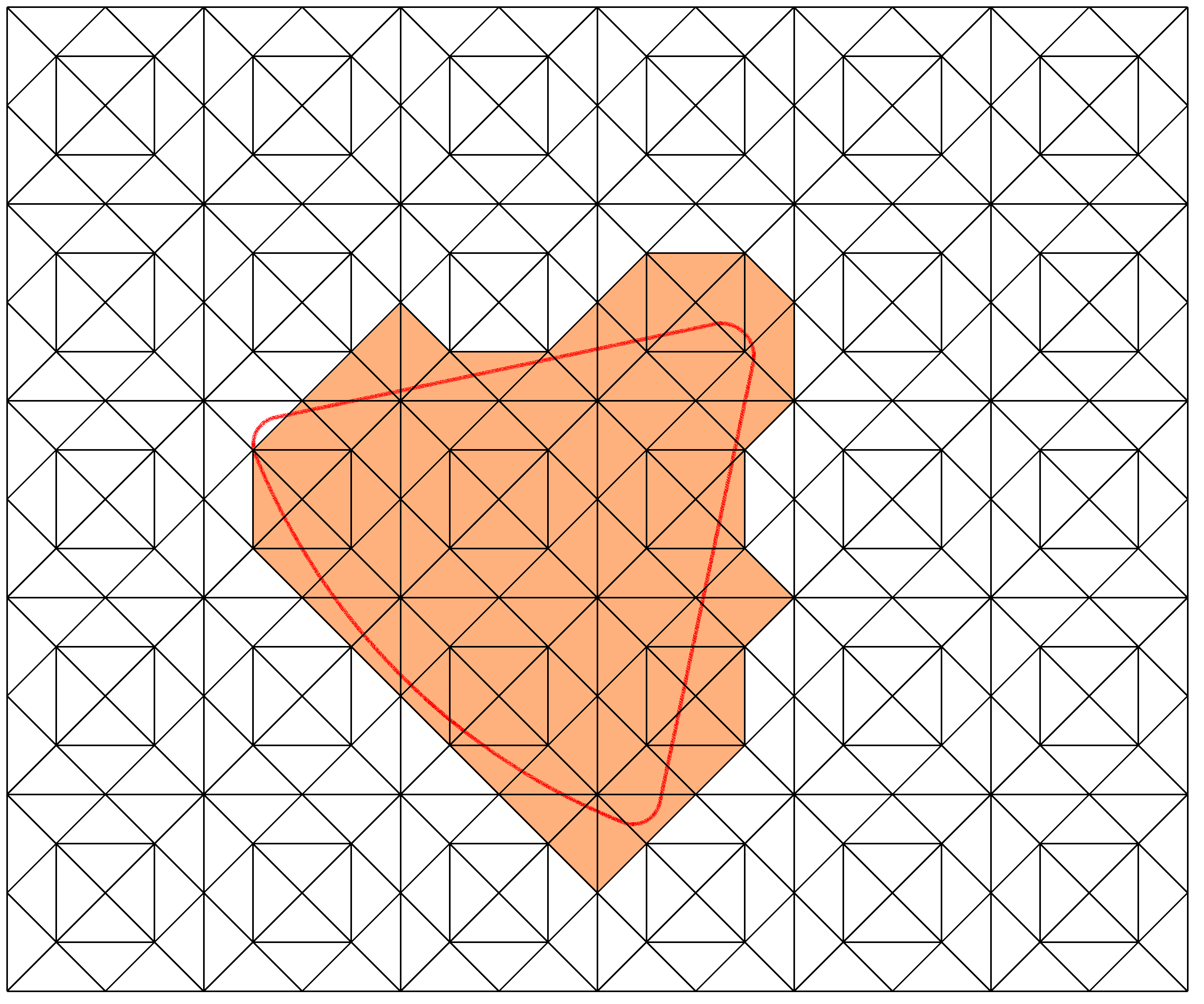} \\ \hspace{-5mm}
\vspace{-5mm} \\
\textrm{(c) $\sigma_B(\bx_B)$ and $\mathcal{M}_{\tt lin}$.} & \textrm{(d) Initial $\mathcal{M}$}\,(p=1). \\ \hspace{-5mm}
\includegraphics[width=0.23\textwidth]{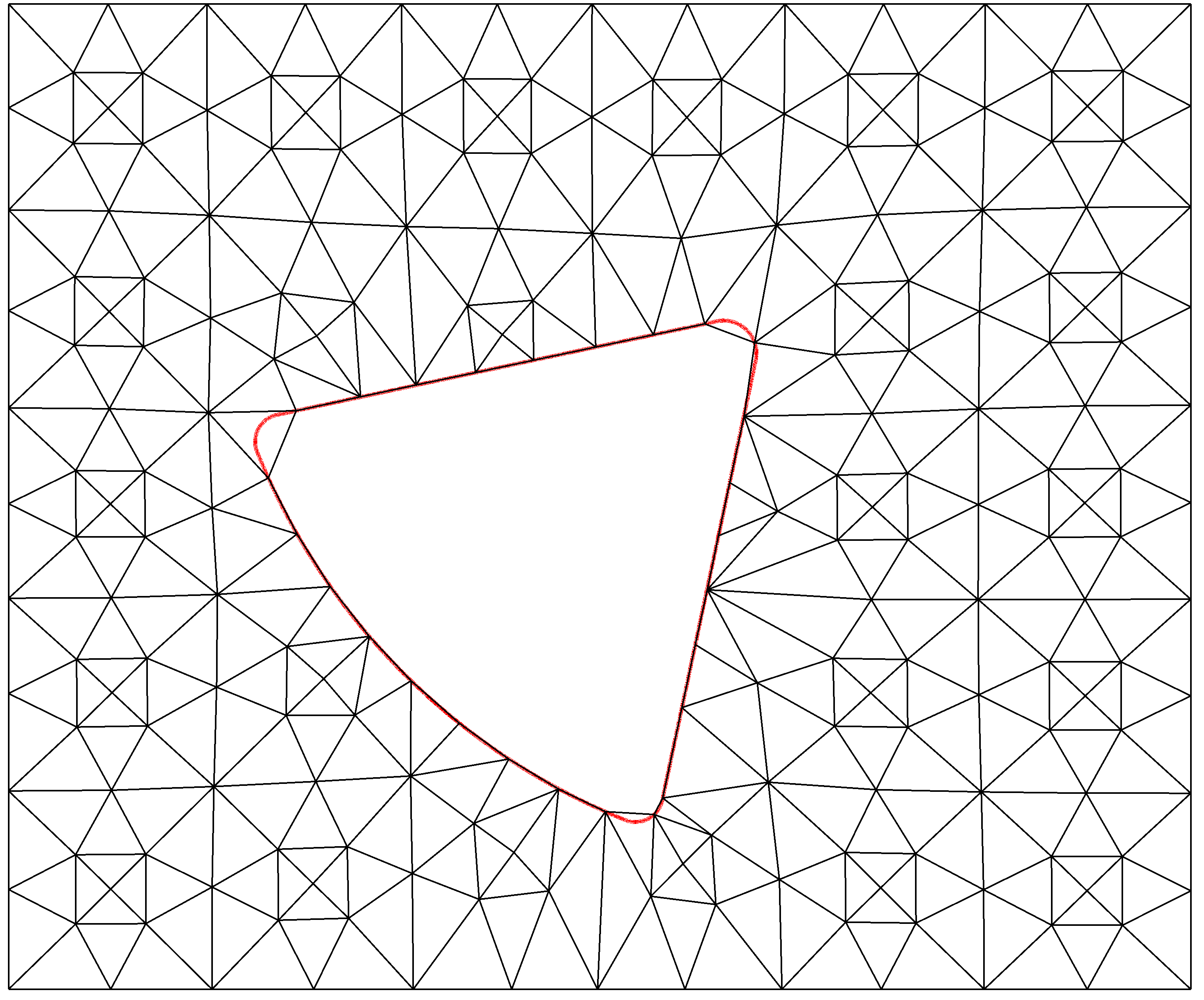} &
\includegraphics[width=0.23\textwidth]{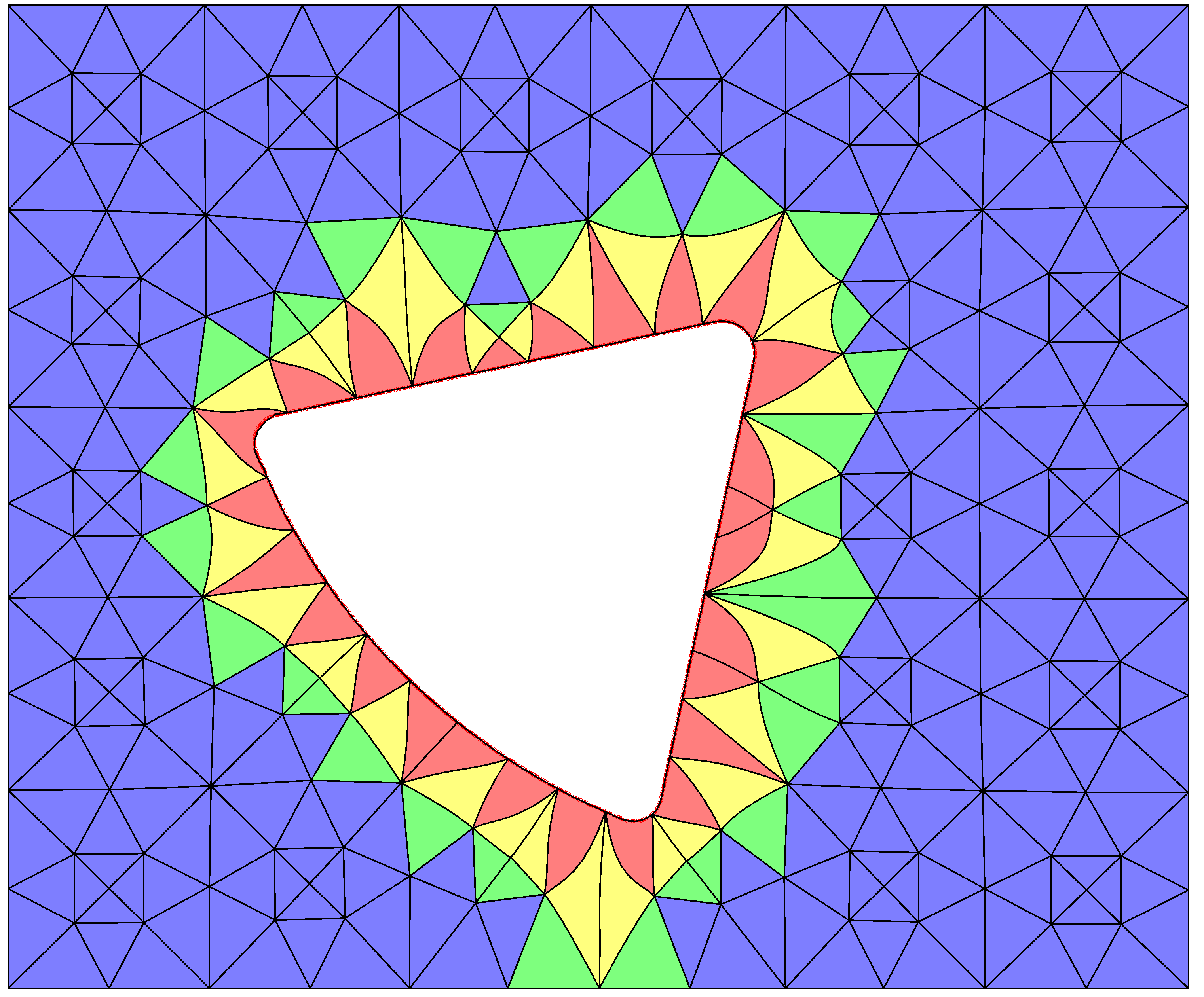} \\ \hspace{-5mm}
\vspace{-5mm} \\
\textrm{\textrm{(e) Optimized $\mathcal{M}$}\,(p=1)}. & \textrm{(f) Optimized $\mathcal{M}$,} \\
& \textrm{$p=4$ around $\mathcal{F}$.} \\
\multicolumn{2}{c}{\includegraphics[width=0.3\textwidth]{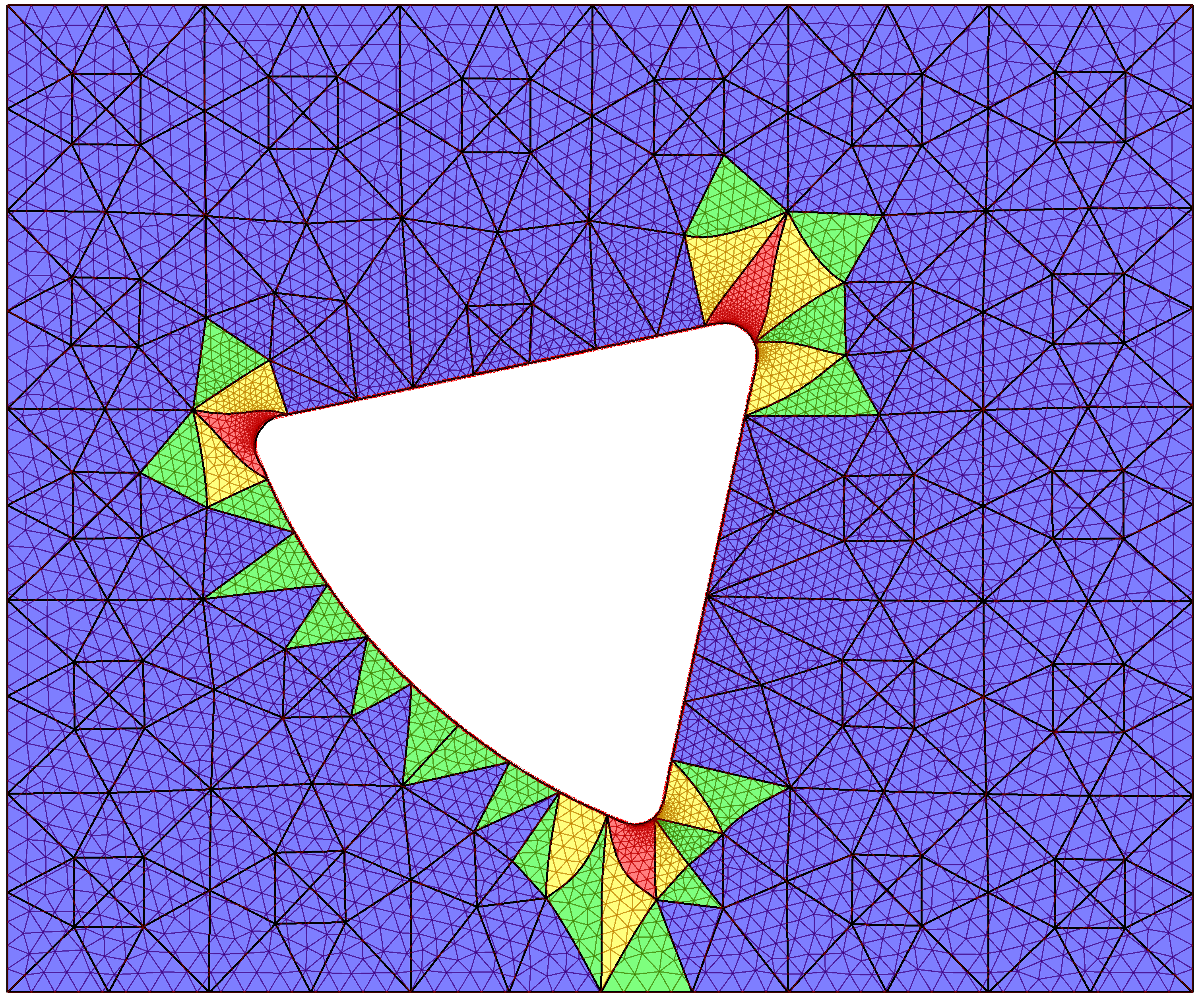}} \\
\multicolumn{2}{c}{\textrm{(g) Mixed-order $\mathcal{M}$, $\sigma_B(\bx_B)=0$, and $\mathcal{M}_{\tt lin}$}} \\
\multicolumn{2}{c}{\includegraphics[width=0.2\textwidth]{figures/legend_rgb_horizontal_1-4}} \\ \hspace{-5mm}
\end{array}$
\end{center}
\vspace{-9mm}
\caption{
Generating a mixed order mesh for the Apollo capsule starting from a dense linear mesh.
The input linear mesh $\mathcal{M}_{\tt lin}$ is used to define a level-set function
$\sigma_B(\bx_B)$ on an adaptively refined background mesh $\mathcal{M}_B$.
A coarse linear mesh $\mathcal{M}$ is then aligned with $\sigma_B(\bx_B)=0$,
and $p$-refined using $p_{max}=4$, $\Delta p=1$, and $\gamma_1=0$.
The $p$-refined mesh is then optimized to align with $\sigma_B(\bx_B)=0$,
and derefined using \eqref{eq_deref_crit_size} with $\beta_3=2\cdot 10^{-4}$
to obtain the final mixed-order mesh.}
\label{fig_apollo}
\vspace{-3mm}
\end{figure}

Figure \ref{fig_apollo}(a) shows a linear mesh ($\mathcal{M}_{\tt lin}$) generated for the 2D Apollo capsule.
Using this linear mesh we construct a background coarse quad mesh
$\mathcal{M}_B$ and adaptively refine
it based on the boundary of $\mathcal{M}_{\tt lin}$. We use findpts (Section \ref{subsec_remap}) to determine which elements of
$\mathcal{M}_B$ intersect with the boundary of $\mathcal{M}_{\tt lin}$.
Figure \ref{fig_apollo}(b) shows the overlap between $\mathcal{M}_B$ and
$\mathcal{M}_{\tt lin}$.
Next, we generate a discrete grid function $\phi_B(\bx_B)$ on
$\mathcal{M}_B$ that indicates whether a given node of $\mathcal{M}_B$ is inside ($\phi_B=+1$) or outside ($\phi_B=-1$) the domain of $\mathcal{M}_{\tt lin}$.
This discrete function is similar to the non-smooth representations that we get for geometric primitives in our CSG-based
approach, as described in the previous section.
The level-set function $\sigma_B$ is then extracted as the
distance function to the zero-isocontour of $\phi_B$.
The error in the zero isocontour of this level-set function $\sigma_B$,
with respect to the true boundary of $\mathcal{M}_{\tt lin}$,
is proportional to the element size of $\mathcal{M}_B$,
and we use 12 levels of AMR for $\mathcal{M}_B$ to decrease this error.
Figure \ref{fig_apollo}(c) shows the level-set function $\sigma_B(\bx_B)$,
its zero isocontour, and the original linear mesh $\mathcal{M}_{\tt lin}$.

With the level-set function $\sigma_B$, we obtain a mixed-order mesh using the approach presented in this work. Starting from a uniform linear triangular mesh, we mark the elements based on whether they are located inside or outside
the domain, see Figure \ref{fig_apollo}(d). The elements determined to be completely outside the domain are removed from the mesh,
and the boundary of the new mesh is aligned to $\sigma_B(\bx_B)=0$.
The optimized linear mesh is shown in Figure \ref{fig_apollo}(e).
Next, we elevate all elements adjacent to the interface to $p_{max}=4$ and use $\Delta p=1$ to ensure that
the maximum different between polynomial order of edge-neighbors is 1.
The optimized mesh is shown in Figure \ref{fig_apollo}(f)
along with the polynomial order of each element.
The final mixed order mesh shown in Figure \ref{fig_apollo}(g) is obtained by using $\beta_3=2\cdot 10^{-4}$.

The total integrated error for the final mixed-order mesh is $10^{-8}$.
The original mesh in Figure \ref{fig_apollo}(d) has 526 elements, which is trimmed
to 390 elements shown in Figure \ref{fig_apollo}(e).
The final optimized mixed-order mesh has 359 linear, 19 quadratic, 9 cubic,
and 4 quartic element.
We note that the maximum local error in the final mixed-order mesh is at the
element with $p=4$ at the left corner.
This is due to the high-curvature of the zero isocontour in that region,
combined with the overall coarse resolution in the mesh.
This is similar to what was observed in the example in the previous section.
In future work, we will seek to address this issue
using $h$-refinement such that local mesh resolution can be enhanced on demand
when $p_{max}$ is not high enough to obtain the desired accuracy.


\section{Summary and Future Work}
\label{sec_concl}

We have presented a novel approach for generating mixed-order meshes
through $rp$-adaptivity in the context of surface fitting.
This approach uses TMOP for ensuring good mesh quality as a mesh is morphed to
align with the surface prescribed as the zero isocontour of a level-set function.
The proposed approach is purely algebraic,
and extends to different element types in 2D and 3D.
We have proposed different possibilities for setting up the mesh
$p$-adaptivity problem, and demonstrated how it can be used to
generate mixed order meshes for problems of practical interest.
We have also identified ways to further improve the effectiveness of the
method, for example, using $h$-refinement to augment local
resolution when needed.
In future work, we will present an automated algorithm based on
$hrp$-adaptivity to robustly obtain accurate mixed order meshes
for different problem types with minimal user input.

\clearpage

\bibliographystyle{siam}
\bibliography{rp}

\end{document}

%% file: tikz-includes.tex
\usepackage{wrapfig}
\usepackage{float}
\usepackage{subfig}
\usepackage{subcaption}
\usepackage[percent]{overpic}
\usepackage{varwidth}
\usepackage{tikz}
\usepackage{pgfplots}
\usepackage{adjustbox}
\usepackage{tikz-layers}

\usepackage{pgfplotstable}
\usepackage{arrayjobx}
\usetikzlibrary{angles, quotes}
\usepackage{tikz-3dplot} 
\usepackage{lmodern}
\usetikzlibrary{shapes.misc}

\tikzset{cross/.style={cross out, draw=red, minimum size=2*(#1-\pgflinewidth), inner sep=0pt, outer sep=0pt},
cross/.default={1pt}}

\usetikzlibrary{patterns}
\usetikzlibrary{plotmarks}
\usetikzlibrary{intersections} 
\usepackage{tkz-euclide}

\usepackage{pgfplots}
\pgfplotsset{compat=1.18}

\pgfplotsset{compat=1.18}
\pgfplotsset{
discard if not/.style 2 args={
    x filter/.append code={
        \edef\tempa{\thisrow{#1}}
        \edef\tempaa{#2}
        \ifx\tempa\tempaa
        \else
            
        \fi
    }
}}

\pgfplotscreateplotcyclelist{my black white}{
  solid, every mark/.append style={solid, fill=gray}, mark=* \\
  dotted, every mark/.append style={solid, fill=gray}, mark=square* \\
  densely dotted, every mark/.append style={solid, fill=gray}, mark=otimes* \\
  loosely dotted, every mark/.append style={solid, fill=gray}, mark=triangle* \\
  dashed, every mark/.append style={solid, fill=gray}, mark=diamond* \\
  loosely dashed, every mark/.append style={solid, fill=gray}, mark=* \\
  densely dashed, every mark/.append style={solid, fill=gray}, mark=square* \\
  dashdotted, every mark/.append style={solid, fill=gray}, mark=otimes* \\
  dashdotdotted, every mark/.append style={solid}, mark=star \\
  densely dashdotted, every mark/.append style={solid, fill=gray}, mark=diamond* \\
}

\pgfplotscreateplotcyclelist{my color}{
  solid, color=red, every mark/.append style={solid, fill=red}, mark=* \\
  solid, color=blue, every mark/.append style={solid, fill=blue}, mark=square* \\
  solid, color=green!50!black, every mark/.append style={solid, fill=green!50!black}, mark=otimes* \\
  solid, color=red!60!white, every mark/.append style={solid, fill=red!60!white}, mark=triangle* \\
  solid, color=orange, every mark/.append style={solid, fill=orange}, mark=diamond* \\
  solid, color=purple, every mark/.append style={solid, fill=purple}, mark=* \\
  solid, every mark/.append style={solid, fill=gray}, mark=square* \\
  solid, every mark/.append style={solid, fill=gray}, mark=otimes* \\
  solid, every mark/.append style={solid}, mark=star \\
  solid, every mark/.append style={solid, fill=gray}, mark=diamond* \\
}

%% file: figures/pconstraint.tex
\begin{tikzpicture}[scale=1]
    \coordinate (0) at (0,0);
    \coordinate (1) at (0.33333333333, 0);
    \coordinate (2) at (0.66666666667,0);
    \coordinate (3) at (1.0,0);
    
    \coordinate (4) at (0.074074,0.33333333333);
    \coordinate (5) at (0.4198, 0.33333333333);
    \coordinate (6) at (0.7654,0.33333333333);
    \coordinate (7) at (1.111,0.33333333333);
    
    \coordinate (8) at (0.1481,0.66666666667);
    \coordinate (9) at (0.4691, 0.66666666667);
    \coordinate (10) at (0.7901,0.66666666667);
    \coordinate (11) at (1.111,0.66666666667);
    
    \coordinate (12) at (0,1);
    \coordinate (13) at (0.33333333333,1);
    \coordinate (14) at (0.66666666667,1);
    \coordinate (15) at (1.0,1);

    \coordinate (23) at (0.009,0.1);
    \coordinate (24) at (0.125,0.5);
    \coordinate (25) at (0.081,0.9);

    \draw [blue!20!white, fill=blue!20!white] plot [smooth] coordinates {(0) (1) (2) (3)} -- (6) -- cycle;
    \draw [blue!20!white, fill=blue!20!white] plot [smooth] coordinates {(3) (7) (11) (15)} -- (6) -- cycle;
    \draw [blue!20!white, fill=blue!20!white] plot [smooth] coordinates {(15) (14) (13) (12)} -- (6) -- cycle;
    \draw [blue!20!white, fill=blue!20!white] plot [smooth] coordinates {(12) (25) (8) (24) (4) (23) (0)} -- (6) -- cycle;
    \draw [blue!20!white, fill=blue!20!white] plot [smooth] coordinates {(6) (10) (9)} -- (5) -- cycle;
    
    \draw[black] plot [smooth cycle] coordinates {(0) (1) (2) (3)} ;
    \draw[black] plot [smooth] coordinates {(15) (14) (13) (12)} ;
    \draw[black] plot [smooth] coordinates {(12) (25) (8) (24) (4) (23) (0)} ;
    
    \coordinate (16) at (1.125,0.5);
    \coordinate (17) at (1.5, 0);
    \coordinate (18) at (2.0, 0);
    
    \coordinate (19) at (1.5625,0.5);
    \coordinate (20) at (2, 0.5);
    
    \coordinate (21) at (1.5, 1);
    \coordinate (22) at (2, 1);
    
    \draw [blue!20!white, fill=blue!20!white] plot [smooth] coordinates {(3) (17) (18)} -- (19) -- cycle;
    \draw [blue!20!white, fill=blue!20!white] plot [smooth] coordinates {(18) (20) (22)} -- (19) -- cycle;
    \draw [blue!20!white, fill=blue!20!white] plot [smooth] coordinates {(22) (21) (15)} -- (19) -- cycle;
    \draw [blue!20!white, fill=blue!20!white] plot [smooth] coordinates {(15) (16) (3)} -- (19) -- cycle;
    
    \draw[black] plot [smooth cycle] coordinates {(3) (17) (18)} ;
    \draw[black] plot [smooth] coordinates {(18) (20) (22)} ;
    \draw[black] plot [smooth] coordinates {(22) (21) (15)} ;

    \draw[black] plot [smooth] coordinates {(3) (7) (11) (15)} ;

    \foreach \i in {0,...,6}
    {
        \draw (\i) node[circle, fill=black, inner sep=1pt] {} ;
    }
    \foreach \i in {8,...,10}
    {
        \draw (\i) node[circle, fill=black, inner sep=1pt] {} ;
    }
    \foreach \i in {12,...,15}
    {
        \draw (\i) node[circle, fill=black, inner sep=1pt] {} ;
    }
    
    \draw (11) node[cross, fill=red, inner sep=1pt] {} ;
    \draw (7) node[cross, fill=red, inner sep=1pt] {} ;
    
    \foreach \i in {16,...,22}
    {
        \draw (\i) node[circle, fill=black, inner sep=1pt] {} ;
    }

    \draw (0.0, -0.4) node[cross, fill=red, inner sep=1pt] {} ;
    \draw (0.6, -0.4) node[scale=0.3] {Constrained DOFs};
    
    \draw (0.425, -0.2) node[scale=0.3] {True DOFs};
    \draw (0.0, -0.2) node[circle, fill=black, inner sep=1pt] {} ;
    
    \draw (0.5, 0.15) node[scale=0.5] {$p=3$};
    \draw (1.5, 0.15) node[scale=0.5] {$p=2$};
    
    \end{tikzpicture}

%% file: figures/error_squircle_uniformp.tex
\begin{tikzpicture}[scale=0.75]
  \begin{axis}[ 
      title={},
      width=4.521in,
      height=3.566in,
      at={(0.758in,0.481in)},
      scale only axis,
      ymode=log,
      ymax=1e-6,
      ymin=1e-14,
      ylabel={Total integrated fitting error ($\int_{\mathcal{F}}\sigma^2(\mathbf{x}_B)$)},
      xlabel={Total DOFs},
      xmajorgrids,
      ymajorgrids,
      legend pos=north east, 
      legend style={legend cell align=left, align=left, draw=white!15!black, font=\huge},
      label style={font=\Large},
      tick label style={font=\large},
      cycle list name=my color,
      xtick distance={5000},
      scaled x ticks=false,
    ]

    \addplot+[line width=2.5pt] table [x=NDofs, y=Error, discard if not={Type}{0}, discard if not={order}{1}] {./figures/2Dsquircle_error.table};
    \addplot+[line width=2.5pt] table [x=NDofs, y=Error, discard if not={Type}{0}, discard if not={order}{2}] {./figures/2Dsquircle_error.table};
    \addplot+[line width=2.5pt] table [x=NDofs, y=Error, discard if not={Type}{0}, discard if not={order}{3}] {./figures/2Dsquircle_error.table};
    \addlegendentry{$p=1$}
    \addlegendentry{$p=2$}
    \addlegendentry{$p=3$}
    \draw[->, black, thick] plot [smooth] coordinates {(5000,1e-9) (8000, 2e-11) (15000,1e-11)};
    \draw (17000,2e-11) node[scale=1.4] {$h-$refinement};

\end{axis}



\end{tikzpicture}

%% file: figures/error_squircle_adaptivep.tex
\begin{tikzpicture}[scale=0.75]
  \begin{axis}[ 
      title={},
      width=4.521in,
      height=3.566in,
      at={(0.758in,0.481in)},
      scale only axis,
      ymode=log,
      ymax=1e-6,
      ymin=1e-14,
      ylabel={Total integrated fitting error ($\int_{\mathcal{F}}\sigma^2(\mathbf{x}_B)$)},
      xlabel={Total DOFs},
      xmajorgrids,
      ymajorgrids,
      legend pos=north east, 
      legend style={legend cell align=left, align=left, draw=white!15!black, font=\huge},
      label style={font=\Large},
      tick label style={font=\large},
      cycle list name=my color,
      xtick distance={5000},
      scaled x ticks=false,
    ]

    \addplot+[line width=2.0pt, mark options={scale=1.5}] table [x=NDofs, y=Error, discard if not={Type}{0}, discard if not={order}{1}] {./figures/2Dsquircle_error.table};
    \addplot+[line width=2.0pt, mark options={scale=1.5}] table [x=NDofs, y=Error, discard if not={Type}{0}, discard if not={order}{2}] {./figures/2Dsquircle_error.table};
    \addplot+[line width=2.0pt, mark options={scale=1.5}] table [x=NDofs, y=Error, discard if not={Type}{0}, discard if not={order}{3}] {./figures/2Dsquircle_error.table};
    \addlegendentry{Uniform $p=1$}
    \addlegendentry{Uniform $p=2$}
    \addlegendentry{Uniform $p=3$}
    \draw[->, black, thick] plot [smooth] coordinates {(5000,1e-9) (8000, 2e-11) (15000,1e-11)};
    \draw (17000,2e-11) node[scale=1.4] {$h-$refinement};


    \addplot+[line width=2.0pt, mark options={scale=1.5},color=orange] table [x=PreNDofs, y=PreError, discard if not={Type}{1},  discard if not={dp}{-1}] {./figures/2Dsquircle_error.table};
    \addlegendentry{$p$-refined A}
    \addplot+[line width=2.0pt, mark options={scale=1.5}, color=purple] table [x=NDofs, y=Error, discard if not={Type}{1},  discard if not={dp}{-1}] {./figures/2Dsquircle_error.table};
    \addlegendentry{$p$-refined B}
    \addplot+[line width=2.0pt, mark options={scale=1.5},color=black] table [x=NDofs, y=Error, discard if not={Type}{1},  discard if not={dp}{1}] {./figures/2Dsquircle_error.table};
    \addlegendentry{$p$-refined C}

    \draw[<->, black, thick] plot [smooth] coordinates {(4000,2e-14) (19000,2e-14)};
    \draw (10000,7e-14) node[scale=1.4] {75\% fewer DOFs};
\end{axis}



\end{tikzpicture}

%% file: figures/error_squircle3d_uniformp.tex
\begin{tikzpicture}[scale=0.75]
  \begin{axis}[ 
      title={},
      width=4.521in,
      height=3.566in,
      at={(0.758in,0.481in)},
      scale only axis,
      ymode=log,
      ymax=1e-6,
      ymin=1e-14,
      ylabel={Total integrated fitting error ($\int_{\mathcal{F}}\sigma^2(\mathbf{x}_B)$)},
      xlabel={Total DOFs ($N$)},
      xmajorgrids,
      ymajorgrids,
      legend pos=north east, 
      legend style={legend cell align=left, align=left, draw=white!15!black, font=\huge},
      label style={font=\Large},
      tick label style={font=\large},
      cycle list name=my color,
      xtick distance={100000},
      scaled x ticks=false,
    ]

    \addplot+[line width=2.5pt, mark options={scale=2.2}, color=black] table [x=NDofs, y=Error, discard if not={Type}{0}] {./figures/3Dsquircle_error.table};
    \addlegendentry{Uniform $p=\{1,2,3\}$}

    \addplot[color=blue, fill=blue, style={thick}, only marks, mark=*, mark options={scale=2.2}]
    coordinates {(35703, 1.7307e-09)};
    \addlegendentry{$p$-refined: $\beta_3 = 10^{-2}$};
    \addplot[color=red, fill=red, style={thick}, only marks, mark=*, mark options={scale=2.2}]
    coordinates {(44667, 1.3504e-11)};
    \addlegendentry{$p$-refined: $\beta_3 = 10^{-3}$};
    \addplot[color=green, fill=green, style={thick}, only marks, mark=*, mark options={scale=2.2}]
    coordinates {(102435, 2.1643e-12)};
    \addlegendentry{$p$-refined: $\beta_3 = 0$};
    \draw[<->, black, thick] plot [smooth] coordinates {(85000,6e-13) (350000,6e-13)};
    \draw (200000,1e-12) node[scale=1.4] {71\% fewer DOFs};
\end{axis}



\end{tikzpicture}